\documentclass[aps,prl,numerical,superscriptaddress,showpacs,floatfix,reprint]{revtex4-1}
\usepackage{amsmath,amssymb,mathrsfs,bm}
\usepackage{url}
\usepackage{hyperref}
\hypersetup{
colorlinks=true,
linkcolor=blue,
anchorcolor =red,
citecolor=blue,
filecolor = red,
urlcolor=blue,
pdfauthor=author}
\usepackage{txfonts}
\usepackage[utf8]{inputenc}
\usepackage{amsmath}
\usepackage{graphicx}
\def\av<#1>{\left\langle\,#1\,\right\rangle}
\def\ev<#1>{\left\langle\,#1\,\right\rangle_{\rm{ev}}}

\begin{document}


\title{Azimuthal Anisotropy Scaling Functions for Identified Particle and Anti-Particle Species \\ across Beam Energies: Insights into Baryon Junction Effects}

\author{ Roy~A.~Lacey}
\email[E-mail: ]{Roy.Lacey@Stonybrook.edu}
\affiliation{Department of Chemistry, 
Stony Brook University, \\
Stony Brook, NY, 11794-3400, USA}
%
%

\date{\today}

\begin{abstract}
Azimuthal anisotropy scaling functions are constructed from species-resolved anisotropy measurements in Pb+Pb ($\sqrt{s_{NN}}=2.76$, $5.02$~TeV) and Au+Au ($\sqrt{s_{NN}}=7.7$--$200$~GeV) collisions to probe baryon transport and medium response at finite baryon chemical potential ($\mu_B$). Within this data-driven framework, meson and baryon anisotropies spanning the collective-flow and quenching regimes collapse onto common scaling curves, enabling quantitative separation of viscous attenuation, radial flow, and hadronic re-scattering. The attenuation scale $k_\beta$ exhibits a non-monotonic beam-energy dependence, coincident with the low-energy rise of hadronic re-scattering, consistent with a temperature-dependent specific shear viscosity featuring a near-minimum near the QCD critical region. A charge-odd baryon--antibaryon separation in the effective radial-flow response is negligible at LHC energies but grows toward lower $\sqrt{s_{NN}}$. This species-uniform, baryon-number–scaling separation across $p,\Lambda,\Xi,\Omega,$ and $d$ disfavors a purely hadronic origin and supports junction-driven net-baryon transport at finite $\mu_B$, enhancing the
experimental visibility of critical dynamics in finite, rapidly evolving systems. Together, these results establish species-resolved scaling functions as a compact and robust tool for constraining baryon stopping, medium opacity, and QGP transport properties.
\end{abstract}


\maketitle

%

Azimuthal anisotropy measurements provide key insight into the quark--gluon
plasma (QGP) formed in high-energy heavy-ion collisions by quantifying the
azimuthal modulation of particle emission relative to the reaction plane.
These anisotropies are commonly expressed through Fourier coefficients $v_n$,
which encode the medium response to the initial collision geometry and its
transport properties~\cite{Ollitrault:1992bk,Voloshin:2008dg}.

The single-particle azimuthal distribution can be written as
\[
E\frac{d^3N}{dp^3}=\frac{1}{2\pi}\frac{d^2N}{p_Tdp_Tdy}
\left(1+2\sum_{n=1}^{\infty}v_n\cos[n(\phi-\Psi_n)]\right),
\]
where $\phi$ is the particle azimuthal angle and $\Psi_n$ denotes the
$n$th-order event plane.
At low transverse momentum ($p_T$), pressure gradients convert the initial
spatial anisotropy into collective flow harmonics that are sensitive to the
equation of state and the specific shear viscosity $\eta/s$
\cite{Heinz:2013th,Gale:2013da}.
At high $p_T$, anisotropy instead reflects the path-length dependence of
partonic energy loss governed by the jet-quenching transport coefficient
$\hat q$~\cite{Bjorken:1982tu,Baier:1996kr,Majumder:2011uk}.
A consistent interpretation of $v_n(p_T)$ across these regimes therefore
requires a framework capable of linking collective expansion and medium
attenuation within a unified description.

Anisotropy scaling functions (ASF) provide a unified, data-driven framework for
constructing scaling representations of azimuthal anisotropy measurements
across collision systems, centralities ($\mathrm{cent}$), and beam energies.
Previous studies demonstrated that ASF collapse charged-hadron $v_2$ and $v_3$
measurements using physically motivated scaling variables that incorporate
initial eccentricities, system size, viscous attenuation, and controlled
viscous corrections~\cite{Majumder:2007zh,Dusling:2009df,Lacey:2024fpb}.
Subsequent extensions to identified particle species established that ASF also
encode sensitivity to radial flow and hadronic re-scattering~\cite{Lacey:2024uky}.
Together, these results show that scaling functions constructed from measured
$v_n(p_T,\mathrm{cent})$ provide simultaneous, data-driven constraints on
viscous attenuation ($\eta/s$), partonic energy loss ($\hat q$), radial-flow
dynamics, and late-stage hadronic interactions.

The present work leverages this established framework to address a more
differential question: how net-baryon transport, and in particular proposed
baryon-junction dynamics, may imprint themselves on azimuthal anisotropy at
finite baryon chemical potential.
Within the ASF framework, distinct scaling coefficients encode separable medium
responses---including viscous attenuation, hadronic re-scattering, and
radial-flow response---whose evolution with beam energy can be tracked
systematically.
This structure provides a natural basis for isolating baryon-number--dependent
effects beyond conventional mass ordering and late-stage hadronic dynamics.

For a fixed system and centrality, the initial eccentricities and bulk transport
properties are common to all particle species.
Species-dependent differences in $v_n(p_T,\mathrm{cent})$ therefore arise from
their coupling to collective radial flow and late-stage hadronic interactions.
Radial flow induces a characteristic mass- and baryon-number--dependent blue
shift, boosting heavier particles to higher $p_T$~\cite{Lacey:2024uky}, while
hadronic re-scattering further modifies $v_n(p_T)$ according to
species-dependent cross sections.
Within ASF, these effects are captured by species-resolved response coefficients
whose beam-energy dependence reflects the evolution of the underlying medium.

At lower beam energies, where the baryon chemical potential $\mu_B$ is large,
baryon junctions have been proposed as a mechanism for enhanced net-baryon
transport that can introduce an additional, baryon-number--dependent
contribution to azimuthal anisotropy.
As topological QCD configurations, junctions transport net baryon number from
beam rapidities toward midrapidity
\cite{Kharzeev:1996sq,Rossi:1977cy,Vance:1998vh,Lewis:2022arg,Magdy:2024dpm},
potentially augmenting baryon stopping beyond conventional string dynamics
\cite{Pratt:2023pee}.
An increased midrapidity net-baryon density steepens local pressure gradients
experienced by baryons, leading to larger flow-driven anisotropies at fixed
$p_T$ for baryons than for antibaryons.
Within a scaling description, such an effect would manifest as a modification
of the effective radial-flow response that follows baryon number.

Within the ASF framework, this picture motivates a targeted test based on
particle--antiparticle comparisons.
A junction-driven contribution is expected to produce an approximately uniform
baryon--antibaryon separation across baryons with $|n_B|=1$
($p,\Lambda,\Xi,\Omega$), while light nuclei such as the deuteron ($|n_B|=2$)
provide an additional lever arm through the expected $|n_B|$ scaling.
As $\sqrt{s_{NN}}$ increases and net-baryon transport diminishes, any such
baryon-number--dependent component should correspondingly weaken, yielding a
clear and testable beam-energy dependence.

By contrast, late-stage hadronic mechanisms---most notably
baryon--antibaryon annihilation---are expected to leave qualitatively different
signatures.
Annihilation acts late in the hadronic phase, preferentially depletes
low-$p_T$ antibaryons, and produces strangeness-ordered distortions with
distinct centrality dependence
\cite{Steinheimer:2012bn,Sun:2017xrx,Xu:2012gf,Zhou:2024cte}.
Such effects correlate with hadronic re-scattering and therefore track the
meson re-scattering response rather than appear as a species-uniform,
baryon-number--dependent modification of radial flow.
Within ASF, these mechanisms are expected to manifest as characteristic
patterns of scaling violation, providing additional discriminatory power.

Motivated by the strong beam-energy dependence of net-baryon transport,
species-resolved anisotropy scaling functions are constructed within the
established ASF framework using measured
$v_n(p_T,\mathrm{cent})$ in Pb+Pb collisions at
$\sqrt{s_{NN}}=2.76$ and $5.02$~TeV and in Au+Au collisions over
$\sqrt{s_{NN}}=7.7$--$200$~GeV.
This construction enables a quantitative, fully data-driven separation of
baryon-number--dependent transport effects from viscous attenuation and
hadronic re-scattering, while simultaneously constraining the equation of
state and QGP transport coefficients.

\begin{figure}
  \includegraphics[width=0.5\textwidth]{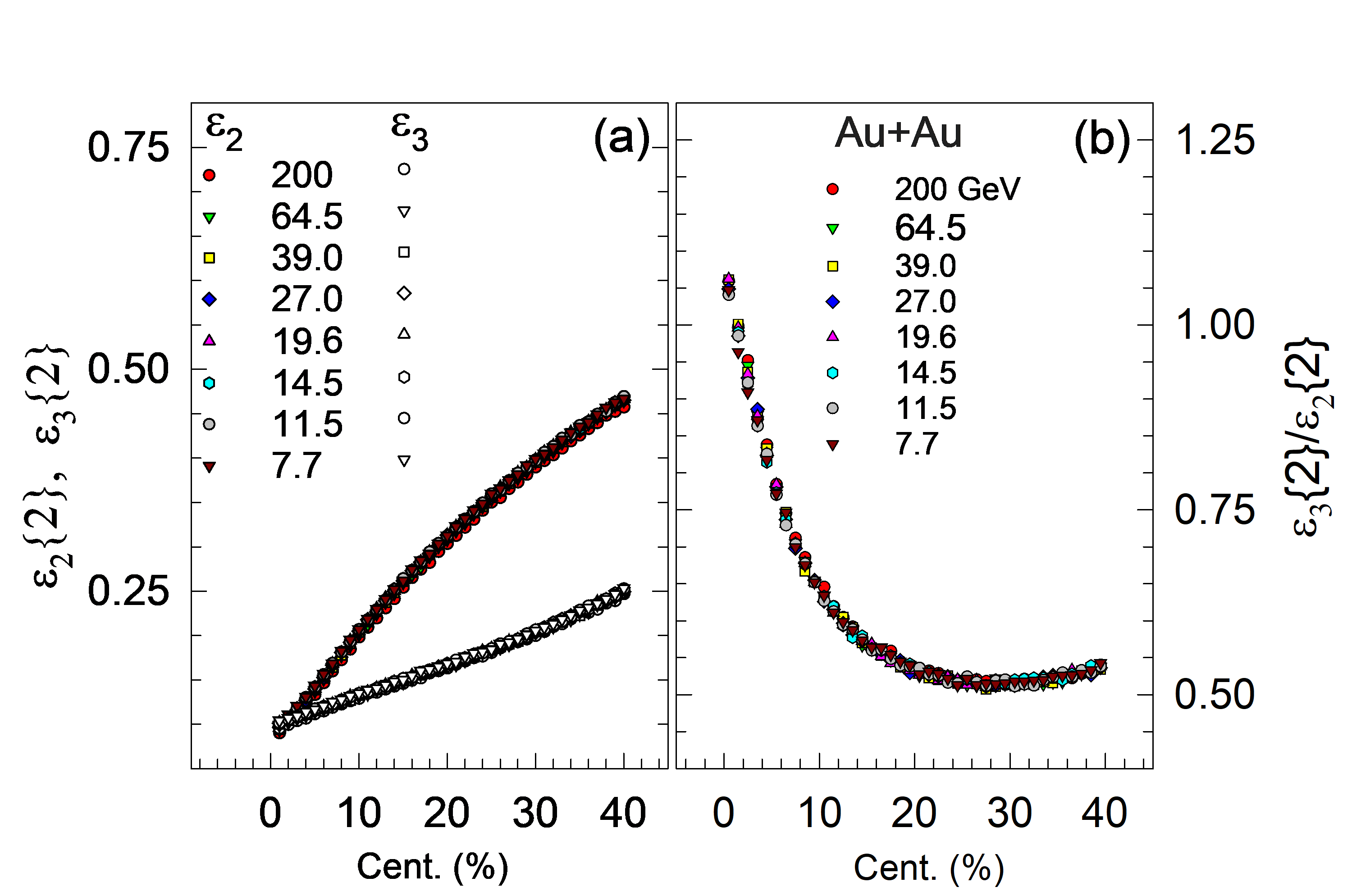}
  \vskip -0.3 cm
\caption{(Color online)
(a) Centrality dependence of the initial-state eccentricities $\varepsilon_2$
and $\varepsilon_3$ for Au+Au collisions at various beam energies.
(b) The ratio $\varepsilon_3/\varepsilon_2$ for Au+Au at the same beam energies,
with the Pb+Pb result at $\sqrt{s_{NN}}=5.02$~TeV included for reference.}
  \label{fig1}
\end{figure}

\begin{figure*}
  \centering
  \includegraphics[width=0.75\textwidth]{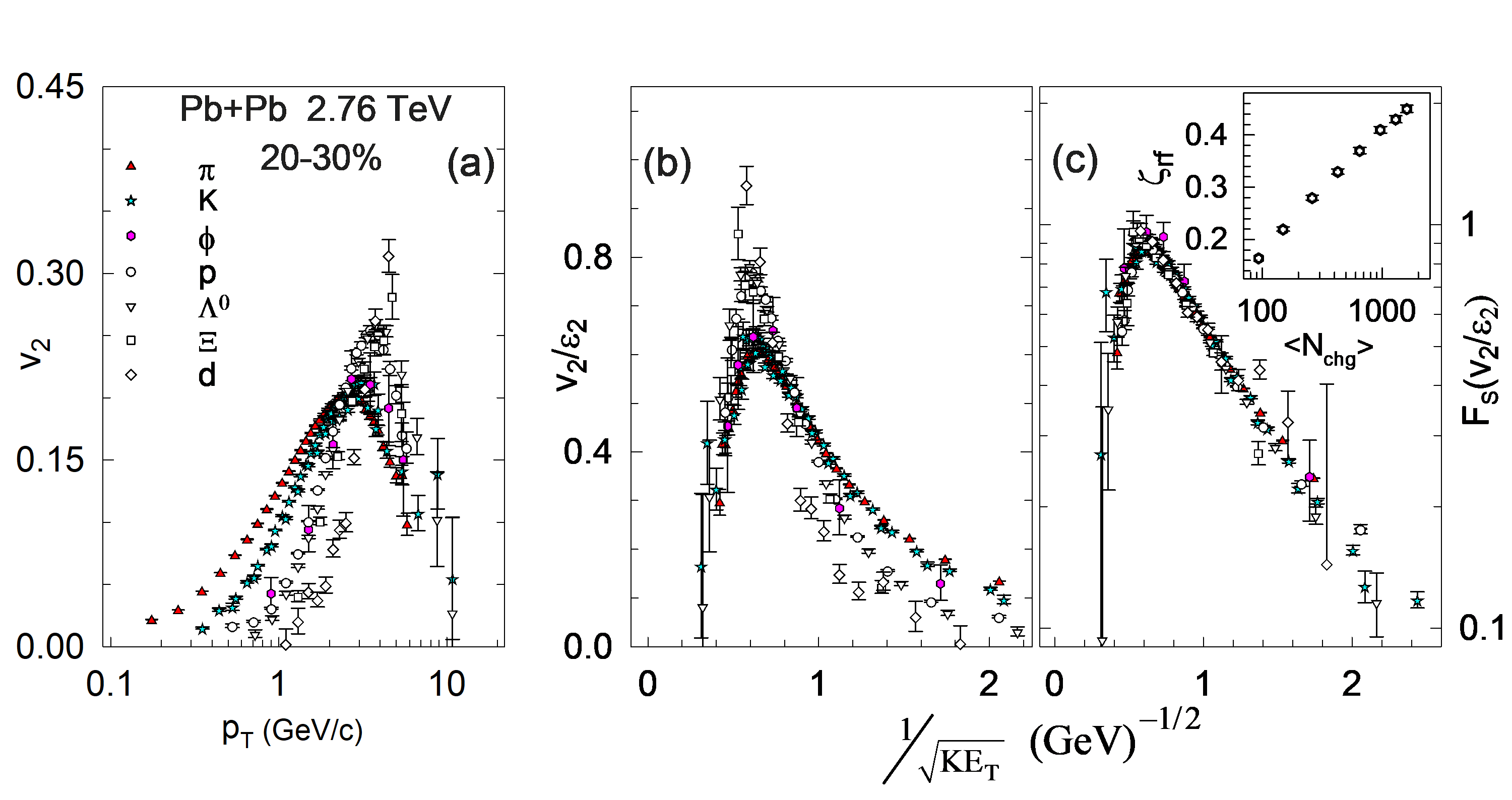}
  \vskip -0.3cm
  \caption{(Color online)
  Panel (a) compares $v_2(p_T)$ for mesons
  ($\pi^{\pm}, K^{\pm}, K^0_S, \phi$) and baryons
  ($d, p, \Lambda^{0}, \Xi^{\pm}, \Omega^{\pm}$), including antiparticles.
  Panel (b) shows the corresponding eccentricity-scaled values
  $v_2/\varepsilon_2$.
  Panel (c) presents the resulting species-resolved scaling function for
  20--30\% central Pb+Pb collisions at
  $\sqrt{s_{NN}}=2.76$~TeV.
  The scaling function is constructed using the kaon reference and the
  species-resolved anisotropy scaling relations described in the text.
  Data are from the ALICE Collaboration
  \cite{Zhu:2019twz,ALICE:2020chv,ALICE:2022zks}.}
  \label{fig2}
\end{figure*}

Species-resolved anisotropy scaling functions are constructed within the
established ASF framework developed for charged hadrons and identified particle
species~\cite{Lacey:2024fpb,Lacey:2024uky}. In this framework, all hadrons share a
common attenuation structure governed by
$\beta = k_\beta \beta_0$, which encodes viscous damping and, through
$\beta \propto \eta/s \propto T^3/\hat q$, carries sensitivity to the jet-quenching
transport coefficient $\hat q$. The geometric response is controlled by the
initial-state eccentricities $\varepsilon_n$ and a transverse-size scale
$\mathbb{R}\propto\langle N_{\rm chg}\rangle_{|\eta|\le0.5}^{1/3}$, which are
common to all particle species for a given system and centrality. Accordingly,
the attenuation parameter $k_\beta$ is taken to be identical for particles and
antiparticles, reflecting the shared viscous and quenching properties of the
medium.

All scaling relations are defined relative to charged kaons in ultra-central
(uc) Pb+Pb collisions at $\sqrt{s_{NN}}=5.02$~TeV, which serve as the common
reference. Kaons provide an optimal baseline owing to their intermediate mass,
small hadronic cross section, and high-precision anisotropy measurements. The
attenuation baseline $\beta_0$ is fixed by this reference, with all
system- and energy-dependent variations absorbed into $k_\beta$.

Species dependence enters exclusively through response coefficients that encode
late-stage dynamics and are extracted directly from the data within the ASF
construction. For mesons, hadronic re-scattering is quantified by
$\zeta_{\rm hs}^{(X)}$, with $\zeta_m^{(X)}=1-\zeta_{\rm hs}^{(X)}$. For baryons,
the collective radial-flow response is encoded through
$\zeta_b^{(X)}=(1-\zeta_{\rm rf}^{(X)})^{|n_B|}$, which explicitly captures
baryon-number dependence. The same functional form is applied to particles and
antiparticles, allowing baryon--antibaryon differences to be isolated through
their extracted response coefficients without introducing additional model
assumptions.

In its full formulation, the species-resolved ASF framework combines
intra-harmonic scaling of $v_2/\varepsilon_2$ with an independent inter-harmonic
mapping between $v_3/\varepsilon_3$ and $v_2/\varepsilon_2$, providing
complementary constraints that enforce a common scaling collapse across
harmonics, systems, and centralities.
In the present analysis, attention is restricted to the $v_2$ sector; however,
the same normalization structure and parameter definitions are retained to
ensure consistency with the full framework.

The meson $v_2$ scaling relation is
\begin{multline}
\frac{v_2(p_T,\mathrm{uc})}{\varepsilon_2(\mathrm{uc})}\,
e^{\frac{2\beta_0}{\mathbb{R}_{\rm uc}}(2+\kappa p_T^2)} =
e^{\alpha\,\frac{2\beta_0}{\mathbb{R}_{\rm uc}}(2+\kappa p_T^2)\zeta_M^{(X)}} \\
\times
\left(\frac{v_2'(p_T)}{\varepsilon_2'}\right)^{\zeta_m^{(X)}}
e^{\frac{2\zeta_m^{(X)}\beta}{\mathbb{R}_{\rm uc}}
\left(\frac{\mathbb{R}_{\rm uc}}{\mathbb{R}'}-1\right)(2+\kappa p_T^2)} ,
\label{eq:v2_scaling_mesons}
\end{multline}
where primes denote the comparison system and centrality.
The normalization exponent
$\zeta_M^{(X)}=\zeta_m^{(X)}+\gamma^{X}_{32}+(1-k_\beta)$
combines the species-dependent meson response with a geometry-only
normalization offset.
The latter,
$
\gamma_{32}^{X}\equiv
\ln\!\left[(\varepsilon_3/\varepsilon_2)_{\rm ref}/
(\varepsilon_3/\varepsilon_2)_{\rm sys}\right]\big|_{X},
$
accounts for inter-system differences in the eccentricity ratio within a fixed
centrality class and enters exclusively through the overall normalization.
By construction, $\gamma_{32}^{X}=0$ for the ultra-central kaon reference system.

The parameter $\alpha$ equals unity for ultra-central collisions, takes an
effective value $\alpha\simeq0.5$ for near-ultra-central selections due to
centrality-bin averaging, and is absorbed into the normalization for non-uc
collisions, where the $p_T$-dependent prefactor on the right-hand side vanishes.

The corresponding baryon $v_2$ scaling relation is
\begin{multline}
\frac{v_2(p_T,\mathrm{uc})}{\varepsilon_2(\mathrm{uc})}\,
e^{\frac{2\beta_0}{\mathbb{R}_{\rm uc}}(2+\kappa p_T^2)} =
e^{(1-\alpha)\,\frac{2\beta_0}{\mathbb{R}_{\rm uc}}(2+\kappa p_T^2)\zeta_B^{(X)}} \\
\times
\left(\frac{v_2'(p_T)}{\varepsilon_2'}\right)^{\zeta_b^{(X)}}
e^{\frac{2\zeta_b^{(X)}\beta}{\mathbb{R}_{\rm uc}}
\left(\frac{\mathbb{R}_{\rm uc}}{\mathbb{R}'}-1\right)(2+\kappa p_T^2)} ,
\label{eq:v2_scaling_baryons}
\end{multline}
where
$\zeta_B^{(X)}=-\zeta_b^{(X)}\!\left(|n_B|/k_\beta-\gamma^{X}_{32}\right)$.
The baryon--antibaryon blue-shift difference is defined as
$\Delta\zeta_{\rm rf}^{(X)}=\zeta_{\rm rf}^{(\bar X)}-\zeta_{\rm rf}^{(X)}$.

Viscous corrections to the thermal distribution are implemented via
$\delta_f=\kappa p_T^2$, with
$\kappa=0.17~(\mathrm{GeV}/c)^{-2}$
\cite{Majumder:2007zh,Dusling:2009df,Liu:2018hjh}.
This correction governs attenuation across the full $p_T$ range; for continuity,
$\delta_f$ is held fixed above
$p_T^{\rm thresh}\!\sim\!4.5$~GeV/$c$, where partonic energy loss dominates.

Taken together, Eqs.~\eqref{eq:v2_scaling_mesons} and
\eqref{eq:v2_scaling_baryons} define a unified, species-resolved ASF framework
that smoothly connects flow-dominated and quenching-dominated regimes and
enables controlled particle--antiparticle tests of baryon-number--dependent
radial-flow response.

Centrality-dependent charged-particle multiplicities used to evaluate the
transverse-size proxy
$\mathbb{R}\propto\langle N_{\text{chg}}\rangle_{|\eta|\le0.5}^{1/3}$
are taken from published measurements
\cite{ALICE:2010mlf,ALICE:2015juo,ALICE:2018cpu,CMS:2019gzk,Lacey:2016hqy}.

Initial-state eccentricities $\varepsilon_n$ are evaluated using the Monte Carlo
quark--Glauber (MC-qGlauber) model~\cite{Liu:2018hjh}, which extends the standard
MC-Glauber framework~\cite{Miller:2007ri,PHOBOS:2006dbo} by incorporating quark
substructure, finite nucleon size, nucleon--quark spatial distributions, and
beam-energy--dependent inelastic cross sections.
Systematic uncertainties on $\varepsilon_n$ are estimated to be 2--3\% from
model-parameter variations.
Figure~\ref{fig1} shows $\varepsilon_{2,3}\{2\}$ and the ratio
$\varepsilon_3/\varepsilon_2$ for Au+Au collisions across beam energies, with
the corresponding $\varepsilon_3/\varepsilon_2$ value for Pb+Pb at
$\sqrt{s_{NN}}=5.02$~TeV included for reference.

The analysis employs $v_2(p_T,\mathrm{cent})$ measurements for identified mesons
($\pi^{\pm}, K^{\pm}, K_S^0, \phi$) and baryons
($^3\mathrm{He}, d, p, \Lambda^{0}, \Xi^{-}, \Omega^{-}$), together with their
antiparticles, from the ALICE, PHENIX, and STAR collaborations.
Scaling functions are constructed for Pb+Pb collisions at
$\sqrt{s_{NN}}=2.76$ and $5.02$~TeV (20--30\% centrality)
\cite{ALICE:2014wao,ALICE:2016cti,ALICE:2017nuf,ALICE:2018lao,
Zhu:2019twz,ALICE:2020chv,ALICE:2022zks},
and for Au+Au collisions over $\sqrt{s_{NN}}=7.7$--$200$~GeV
(10--40\% centrality)
\cite{STAR:2013ayu,PHENIX:2014uik,STAR:2015gge,STAR:2022ncy}.
The ultra-central charged-kaon reference in Pb+Pb at
$\sqrt{s_{NN}}=5.02$~TeV fixes the attenuation baseline $\beta_0$; within each
centrality class, charged kaons define the species baseline entering
Eqs.~\eqref{eq:v2_scaling_mesons} and~\eqref{eq:v2_scaling_baryons}.

To facilitate comparisons among species with different rest masses, the
transverse kinetic energy ${\rm KE}_T=m_T-m_0$, with
$m_T=\sqrt{p_T^2+m_0^2}$, is used as the scaling variable.
This choice suppresses trivial kinematic mass effects and emphasizes collective
dynamics, enabling consistent scaling comparisons between mesons and baryons.

Figure~\ref{fig2} illustrates the species-resolved anisotropy scaling procedure
for 20--30\% central Pb+Pb collisions at $\sqrt{s_{NN}}=2.76$~TeV.
Particle and antiparticle results are averaged, as their differences are within
experimental uncertainties at this energy.
Panel~(a) shows the measured $v_2(p_T)$, which exhibits clear species-dependent
structure.
\begin{figure*}
  \includegraphics[width=0.70\textwidth]{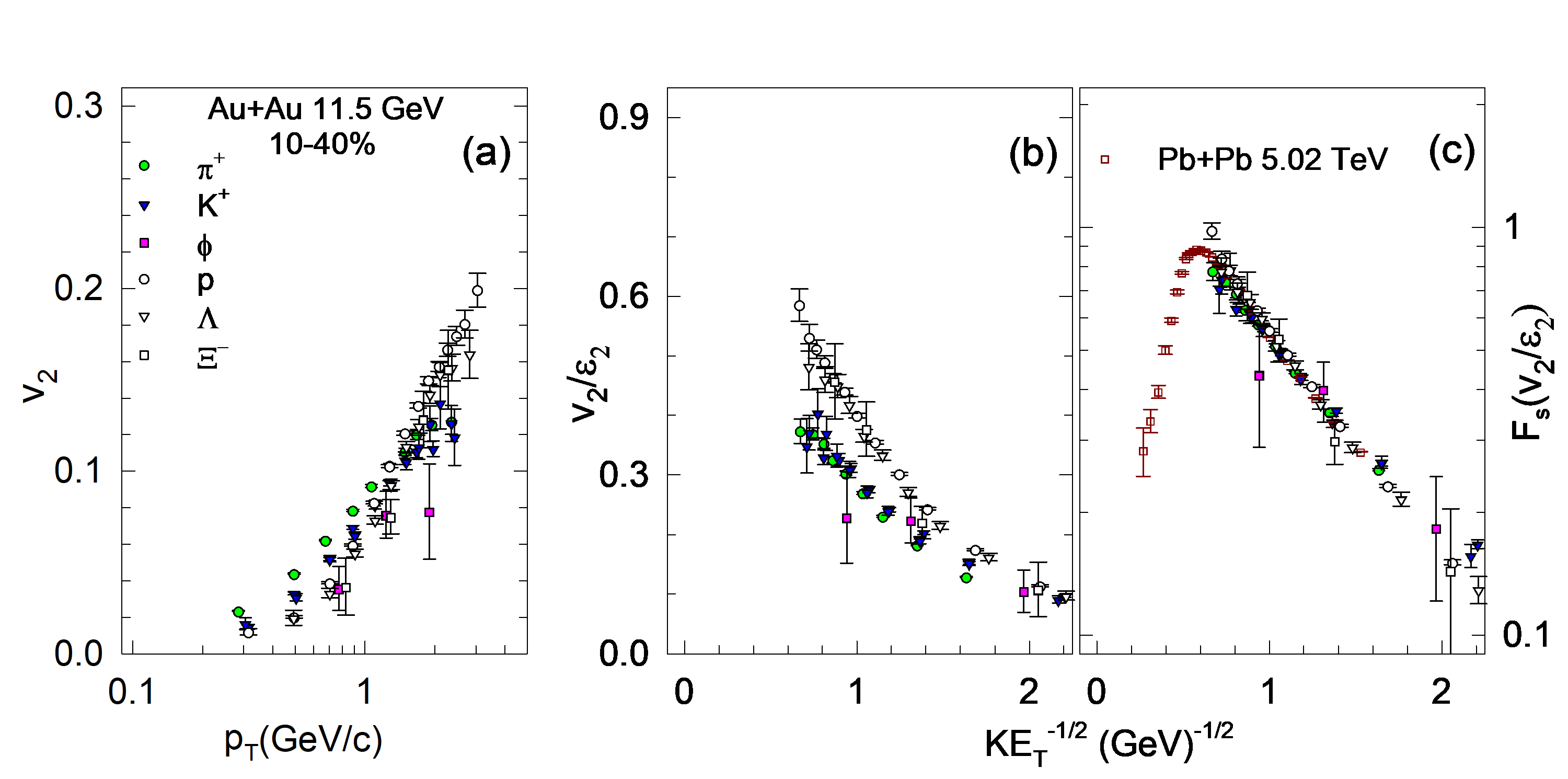}\\
  \vskip -0.1 cm
  \includegraphics[width=0.70\textwidth]{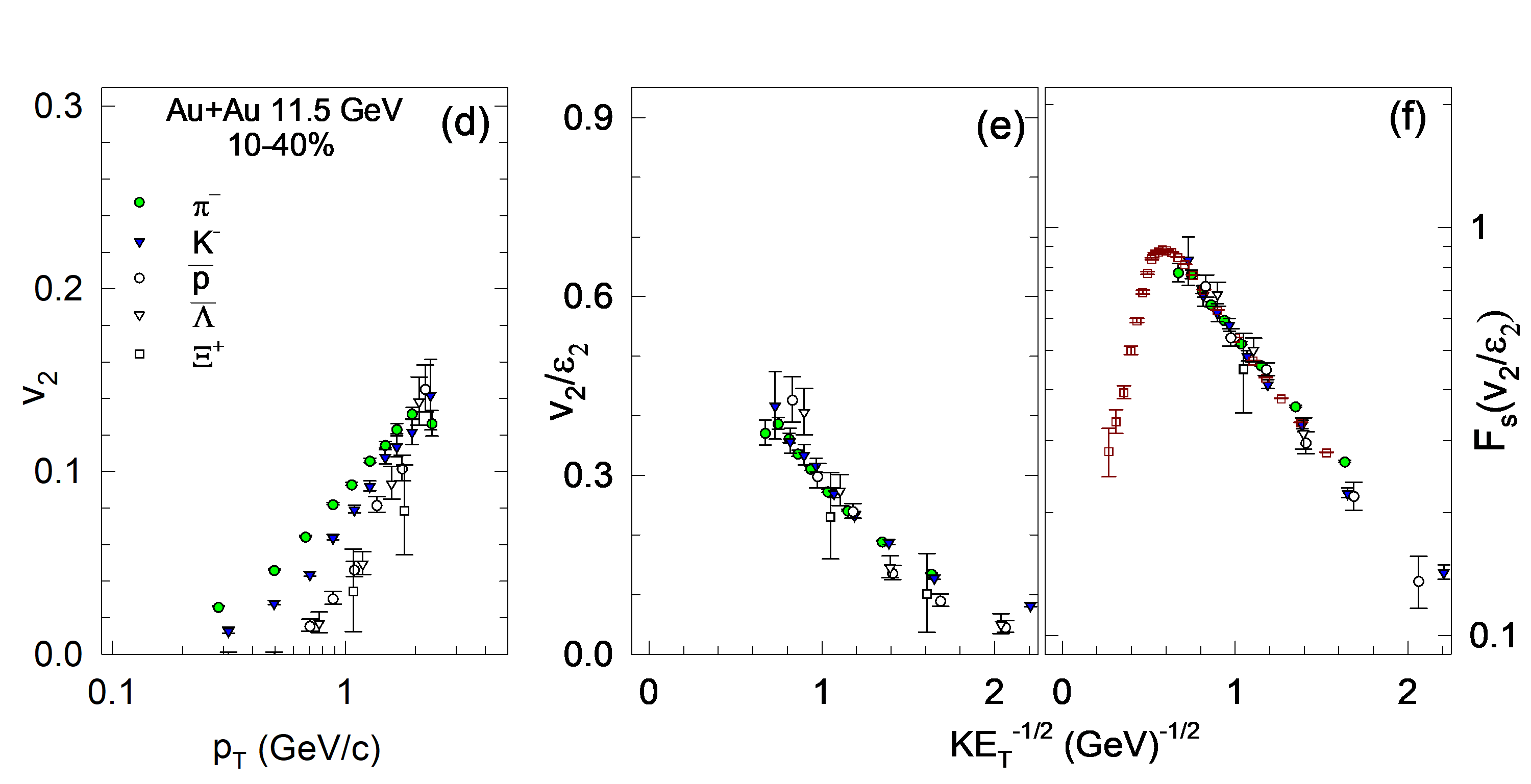}
  \vskip -0.3 cm
 \caption{(Color online) Panels (a--c): $v_2(p_T)$, $v_2/\varepsilon_2$, and the scaling function for identified particles in 10--40\% central Au+Au at $\sqrt{s_{NN}}=11.5$~GeV. Panels (d--f): corresponding antiparticle results. The 20--30\% Pb+Pb baseline at 5.02~TeV is included for comparison. Data from STAR~\cite{STAR:2013ayu} and ALICE~\cite{Zhu:2019twz,ALICE:2022zks}.}
  \label{fig3}
\end{figure*}

As shown in panel~(b), eccentricity scaling, $v_2/\varepsilon_2$, combined with
${\rm KE}_T$ scaling and plotted versus $1/\sqrt{{\rm KE}_T}$
\cite{Dokshitzer:2001zm,Lacey:2010fe}, substantially reduces the meson--baryon
splitting but does not eliminate it.
After this partial collapse, all baryons exhibit a residual blue shift:
single baryons ($|n_B|=1$) cluster with a common offset, while the deuteron
($|n_B|=2$) shows a markedly larger shift, consistent with baryon-number scaling.
Among mesons, pions and kaons track each other closely, whereas the $\phi$ meson
exhibits a small residual deviation at large $1/\sqrt{{\rm KE}_T}$, consistent
with its larger mass and minimal hadronic re-scattering.

Panel~(c) presents the final species-resolved scaling function
$\mathrm{F_S}(v_2/\varepsilon_2)$ obtained using the full ASF relations.
With the parameters summarized in Table~I, all hadron species collapse onto a
single scaling curve.
The negligible meson re-scattering parameter $\zeta_{\rm hs}$ confirms minimal
hadronic modification in this system, while the successful scaling of the
deuteron provides a stringent validation of the baryon-number dependence encoded
in $\zeta_b=(1-\zeta_{\rm rf})^{|n_B|}$.

The robustness of the scaling was verified across the full centrality range for
Pb+Pb at 2.76~TeV.
In all cases, a single baryon radial-flow parameter $\zeta_{\rm rf}$ describes
all baryon species without additional species-dependent tuning.
The extracted $\zeta_{\rm rf}$ values, shown in the Fig.~\ref{fig2} inset as a
function of the mean charged-particle multiplicity
$\langle N_{\rm chg} \rangle_{|\eta|\le0.5}$, increase monotonically toward more
central collisions, following an approximately logarithmic trend consistent
with stronger radial flow at higher energy density.

\begin{figure*}
  \includegraphics[width=0.70\textwidth]{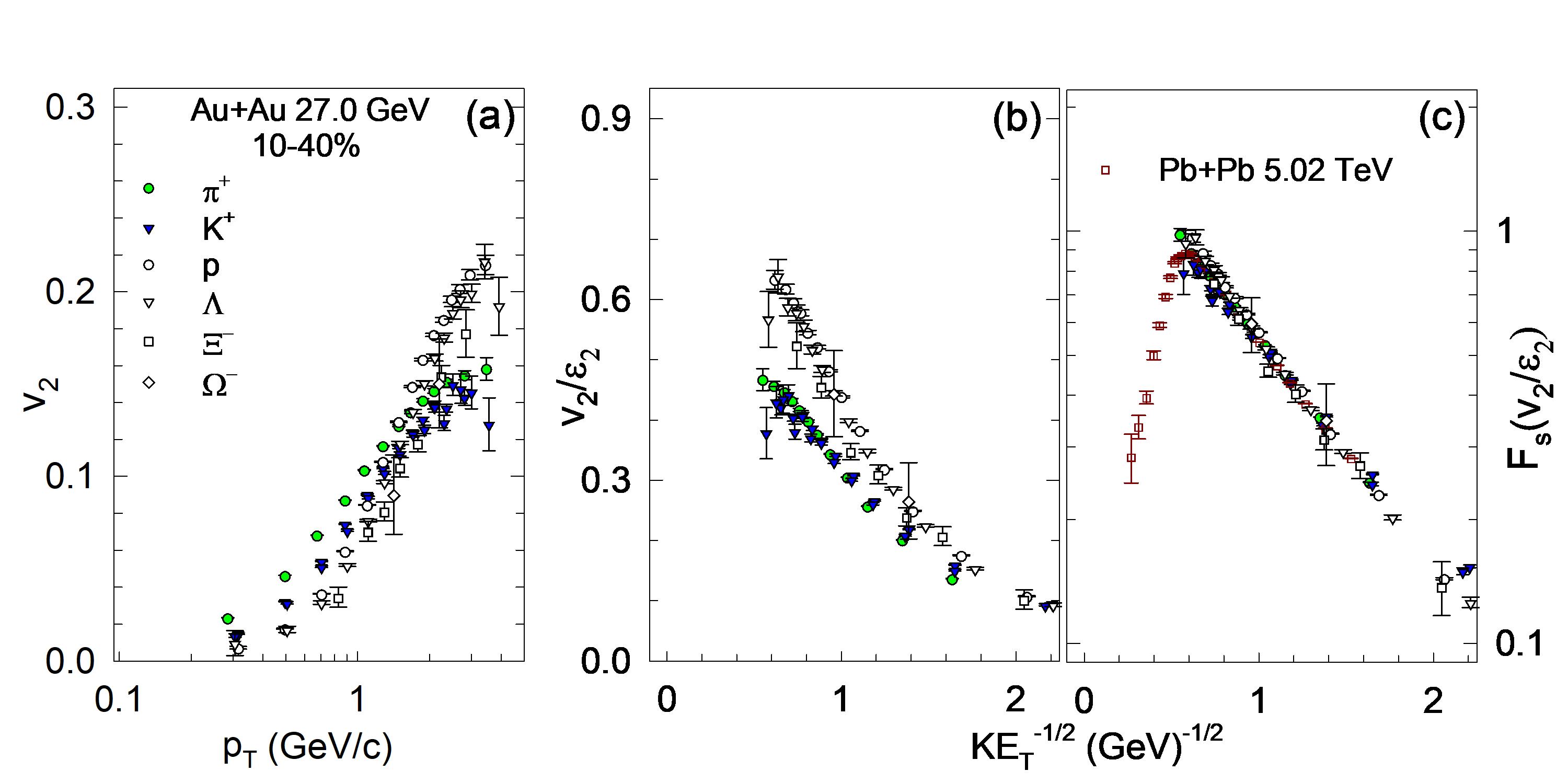}\\
  \vskip -0.1 cm
  \includegraphics[width=0.70\textwidth]{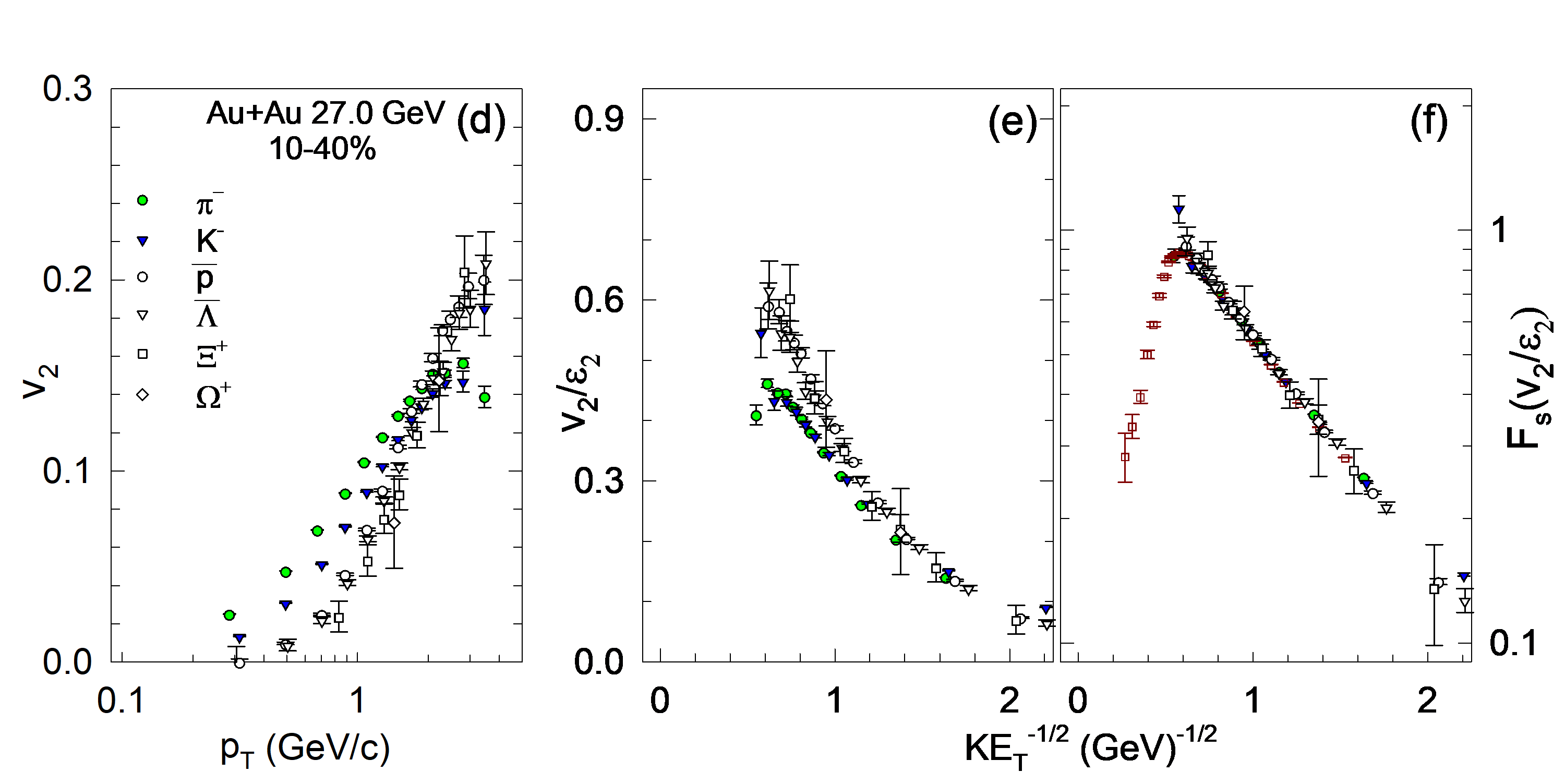}
  \vskip -0.3 cm
  \caption{(Color Online) Same as Fig.~\ref{fig3} but for $\sqrt{s_{NN}} = 27.0$~GeV.}
  \label{fig4}
\end{figure*}

Figures~\ref{fig3} and \ref{fig4} extend the species-resolved scaling analysis to
10--40\% central Au+Au collisions at $\sqrt{s_{NN}}=11.5$ and $27.0$~GeV,
respectively.
Results are shown separately for particles [panels~(a--c)] and antiparticles
[panels~(d--f)].
The measured $v_2(p_T)$ distributions [panels~(a) and (d)] exhibit pronounced
species-dependent structure and, at the lower beam energy, sizable
baryon--antibaryon differences.

After eccentricity scaling, $v_2/\varepsilon_2$, and plotting versus
$1/\sqrt{{\rm KE}_T}$ [panels~(b) and (e)], much of the meson--meson and
baryon--baryon separation is reduced, organizing the data into distinct meson
and baryon bands within each charge sector.
Residual species-dependent differences persist, particularly among mesons at
large $1/\sqrt{{\rm KE}_T}$ (low ${\rm KE}_T$), where mass-dependent radial flow
and hadronic re-scattering are expected to be most influential.
A clear baryon--antibaryon offset remains in the baryon sector, with
antibaryons consistently more blue-shifted than baryons.
The magnitude of this charge-odd separation is largest at
$\sqrt{s_{NN}}=11.5$~GeV and reduced at $27.0$~GeV.

The final species-resolved scaling functions obtained using the full ASF
relations are shown in panels~(c) and (f).
Within each charge sector, the data collapse onto a common scaling curve to good
approximation, with noticeably improved scaling fidelity at
$\sqrt{s_{NN}}=27.0$~GeV compared with $11.5$~GeV.
For reference, the 20--30\% Pb+Pb result at $\sqrt{s_{NN}}=5.02$~TeV is overlaid
as a high-energy baseline.
A comparison of panels~(c) and (f) further indicates that scaling is more robust
for particles than for antiparticles at the lower beam energy, signaling the
onset of controlled, energy-dependent scaling deviations.

The parameters extracted from the scaling analyses in
Figs.~\ref{fig2}--\ref{fig4} are summarized in Table~I.
Relative to the Pb+Pb baseline, Au+Au collisions show a systematically reduced
attenuation scale $k_\beta$, enhanced hadronic re-scattering, and an increasingly
pronounced charge-odd radial-flow response as $\sqrt{s_{NN}}$ decreases.
At $\sqrt{s_{NN}}=200$~GeV, charge-odd effects are small and particle and
antiparticle data are well described by a single radial-flow parameter
$\zeta_{\rm rf}$.
Toward lower beam energies, the separation between
$\zeta_{\rm rf}^{(B)}$ and $\zeta_{\rm rf}^{(\bar B)}$ grows steadily,
signaling increasing net-baryon transport, while remaining species-uniform
across $p,\Lambda,\Xi$, and $\Omega$.
\begin{table}[t]
\centering
\caption{Summary of scaling parameters used/extracted in the species-resolved anisotropy
scaling functions for Figs.~2--4. The attenuation baseline $\beta_0$ is fixed by
ultra-central Pb+Pb at $\sqrt{s_{NN}}=5.02$~TeV.}
\begin{tabular}{lcccccccc}
\hline
System & $\sqrt{s_{NN}}$ (TeV) & Cent. &
$k_\beta$ & $\zeta_{\rm hs}$ &
$\zeta_{\rm rf}^{(B)}$ & $\zeta_{\rm rf}^{(\bar B)}$ &
$\tfrac12\Delta\zeta_{\rm rf}$ & $\gamma_{32}$ \\
\hline
Pb+Pb & 5.02 & 20--30\% & 1.00 & 0.00 & 0.40 & 0.40 & 0.00 & 0.00 \\
Pb+Pb & 2.76 & 20--30\% & 0.95 & 0.00 & 0.38 & 0.38 & 0.00 & 0.00 \\
Au+Au & .200  & 10--40\% & 0.630 & 0.080 & 0.25 & 0.30 & 0.028 & 0.10 \\
Au+Au & .027 & 10--40\% & 0.624 & 0.080 & 0.16 & 0.245 & 0.045 & 0.10 \\
Au+Au & .0115 & 10--40\% & 0.70 & 0.095 & 0.11 & 0.25 & 0.07 & 0.10 \\
\hline
\end{tabular}
\label{tab:ASF_params}
\end{table}

\begin{figure}[t]
    \centering
    \includegraphics[clip,width=1.0\linewidth]{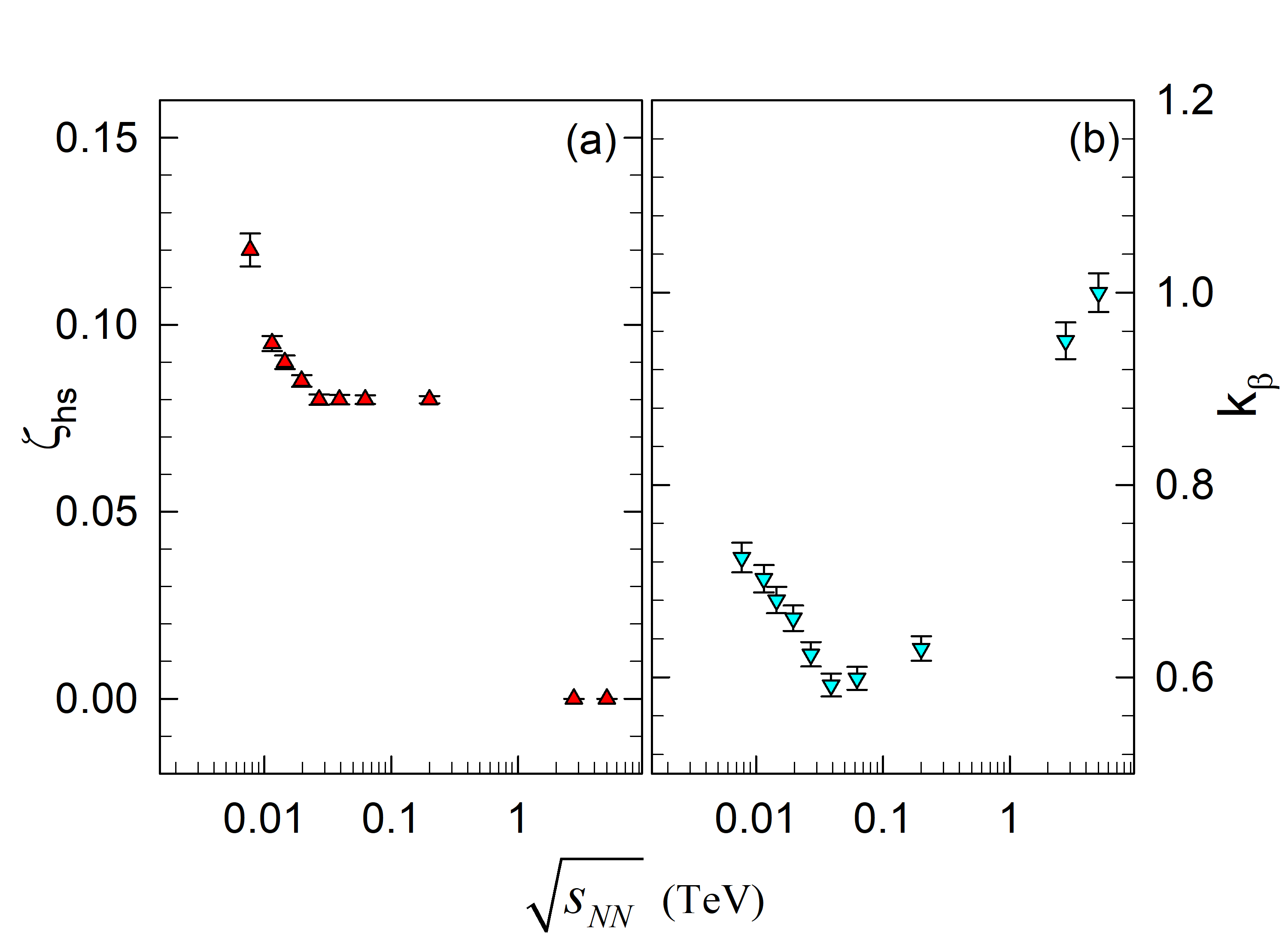}
    \vskip -0.3 cm
   \caption{(Color online)
Beam-energy dependence of (a) the meson re-scattering parameter $\zeta_{\rm hs}$
and (b) the attenuation scale $k_{\beta}\equiv\beta/\beta_{0}$, which serves as a
proxy for the effective specific shear viscosity.
The reference value $\beta_{0}$ is fixed by ultra-central charged kaons in Pb+Pb
at $\sqrt{s_{NN}}=5.02$~TeV, corresponding to $k_{\beta}=1$.
Results are shown for 20--30\% Pb+Pb at $\sqrt{s_{NN}}=2.76$ and 5.02~TeV and
10--40\% Au+Au over $\sqrt{s_{NN}}=7.7$--200~GeV.
Data are from the ALICE, PHENIX, and STAR collaborations
\cite{ALICE:2014wao,ALICE:2016cti,ALICE:2017nuf,ALICE:2018lao,
Zhu:2019twz,ALICE:2020chv,ALICE:2022zks,STAR:2013ayu,
PHENIX:2014uik,STAR:2015gge,STAR:2022ncy}.}
\label{fig5}
\end{figure}

Scaling functions were evaluated across the full beam-energy range to track the
evolution of the key response coefficients.
Excellent scaling fidelity is observed from LHC energies down to
$\sqrt{s_{NN}}\sim 14.5$~GeV, demonstrating that the species-resolved ASF framework
provides a robust description of azimuthal anisotropy over a broad range of
system conditions.
At the lowest BES energies, indications of scaling degradation begin to emerge;
in particular, at $\sqrt{s_{NN}}=7.7$~GeV, deviations are most pronounced for
antibaryons, especially at low ${\rm KE}_T$, signaling an increasing influence
of late-stage hadronic dynamics.

Figure~\ref{fig5} summarizes the beam-energy evolution of the meson
re-scattering parameter $\zeta_{\rm hs}$ [panel~(a)] and the attenuation scale
$k_\beta$ [panel~(b)], which serves as a proxy for the effective specific shear
viscosity.
The $\zeta_{\rm hs}$ values shown are extracted from particle data; over the
beam-energy interval where scaling fidelity is excellent, particle and
antiparticle extractions yield consistent $\zeta_{\rm hs}$ within uncertainties,
indicating that hadronic re-scattering enters predominantly as a charge-even
medium response.

As shown in Fig.~\ref{fig5}(a), $\zeta_{\rm hs}$ decreases with increasing
$\sqrt{s_{NN}}$, is consistent with a constant over $27$--$200$~GeV within
uncertainties, and approaches zero at LHC energies.
This behavior reflects a balance between enhanced hadron density at lower beam
energy and reduced hadronic lifetimes and effective path lengths associated with
more rapid longitudinal and transverse expansion.

In contrast, Fig.~\ref{fig5}(b) shows that $k_\beta$ exhibits a non-monotonic
dependence on $\sqrt{s_{NN}}$, decreasing from 5.02~TeV to $\sim$39~GeV before
rising again at the lowest BES energies.
The correlated evolution of $k_\beta$ and $\zeta_{\rm hs}$ toward low
$\sqrt{s_{NN}}$ indicates a growing influence of the hadronic phase on viscous
attenuation, qualitatively consistent with expectations for a temperature-
dependent $\eta/s$ featuring a near-minimum in the vicinity of the QCD phase
transition~\cite{Csernai:2006zz,Lacey:2006bc}.
In a finite, rapidly evolving system, such behavior is naturally smoothed by
finite-size and finite-time effects, in contrast to the sharp non-monotonic
structure anticipated in the thermodynamic limit.

\begin{figure}[t]
    \centering
    \includegraphics[clip,width=1.0\linewidth]{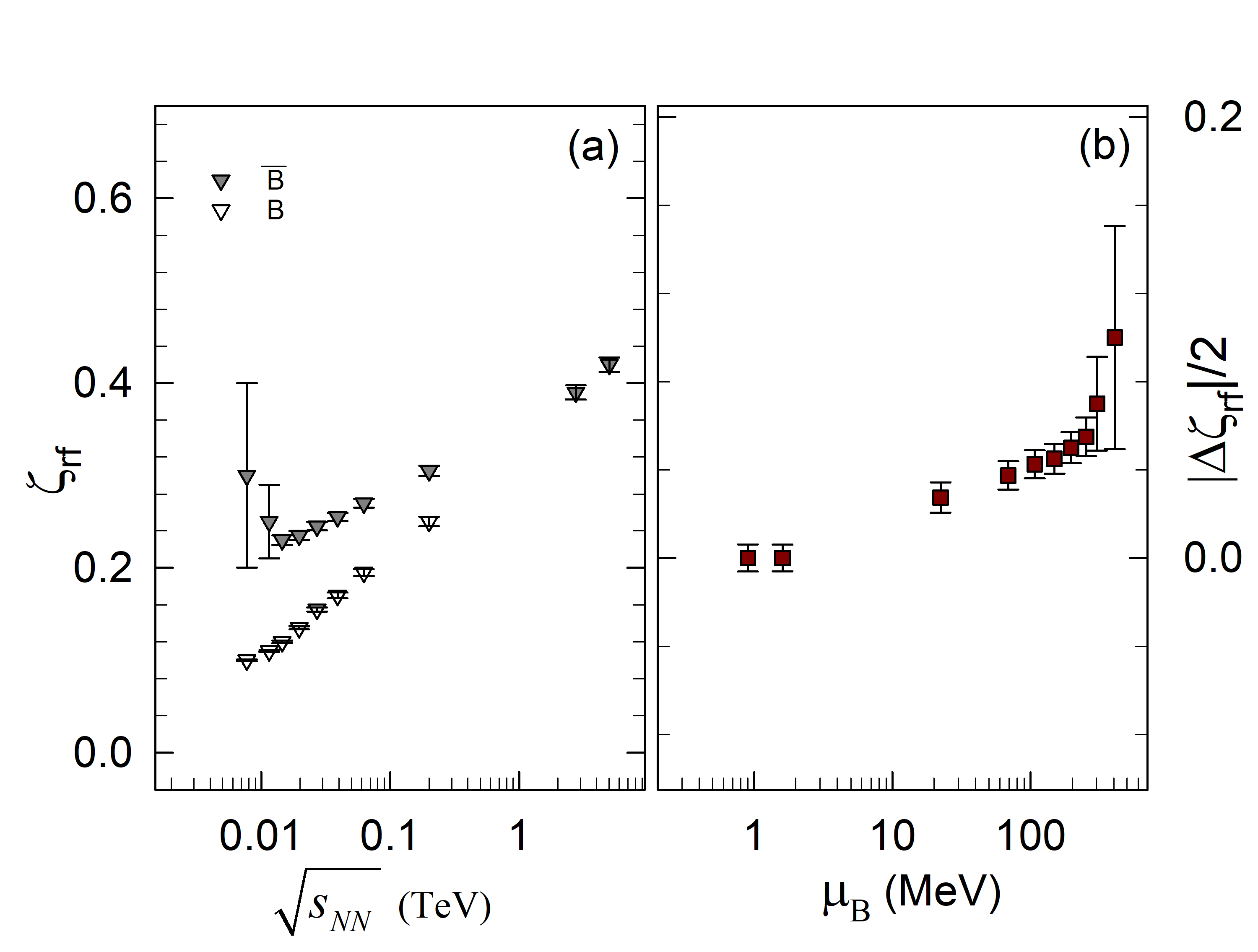}
    \vskip -0.3 cm
    \caption{(Color online)
Beam-energy dependence of (a) the baryon and antibaryon radial-flow response
parameters $\zeta_{\rm rf}^{(B)}$ and $\zeta_{\rm rf}^{(\bar B)}$, shown as a
function of $\sqrt{s_{NN}}$, and (b) half the charge-odd separation
$\tfrac12\Delta\zeta_{\rm rf}$, with
$\Delta\zeta_{\rm rf}\equiv\zeta_{\rm rf}^{(\bar B)}-\zeta_{\rm rf}^{(B)}$,
shown as a function of the baryon chemical potential $\mu_B$.
Results are obtained from species-resolved scaling for 20--30\% Pb+Pb
collisions at $\sqrt{s_{NN}}=2.76$ and 5.02~TeV and for 10--40\% Au+Au
collisions over $\sqrt{s_{NN}}=7.7$--200~GeV.
Data are from the ALICE, PHENIX, and STAR collaborations
\cite{ALICE:2014wao,ALICE:2016cti,ALICE:2017nuf,ALICE:2018lao,Zhu:2019twz,
ALICE:2020chv,ALICE:2022zks,STAR:2013ayu,PHENIX:2014uik,STAR:2015gge,
STAR:2022ncy}.}
\label{fig6}
\end{figure}

Figure~\ref{fig6}(a) shows the beam-energy dependence of the baryon and
antibaryon radial-flow response parameters.
From top LHC energies down to
$\sqrt{s_{NN}}\!\sim\!14.5$~GeV, both $\zeta_{\rm rf}^{(B)}$ and
$\zeta_{\rm rf}^{(\bar B)}$ follow smooth, approximately logarithmic trends with
$\sqrt{s_{NN}}$, with antibaryons consistently exhibiting a larger response than
baryons at the same energy and centrality.
Over this interval, all measured single-baryon species
($p,\Lambda,\Xi,\Omega$) are described by a common $\zeta_{\rm rf}^{(B)}$, and all
corresponding antibaryons by a common $\zeta_{\rm rf}^{(\bar B)}$, demonstrating
a species-uniform, charge-odd modification of the radial-flow response that
tracks baryon number rather than mass or hadronic cross section.

Such behavior is naturally compatible with a junction-mediated contribution to net-baryon transport. The persistence of this charge-odd separation across species and over a broad
beam-energy range supports an early-time, medium-driven origin rather than a
late-stage hadronic effect.

At lower beam energies, beginning around
$\sqrt{s_{NN}}\!\approx\!11.5$~GeV, the scaling fidelity for antibaryons begins to
degrade, most noticeably at low ${\rm KE}_T$.
While $\zeta_{\rm rf}^{(B)}$ continues to evolve smoothly with decreasing
$\sqrt{s_{NN}}$, the antibaryon response exhibits increased scatter and
sensitivity to additional dynamics.
In this regime, late-stage hadronic processes—such as antibaryon annihilation
and associated \emph{flow filtering}—can bias the effective antibaryon response
and partially obscure the underlying transport-driven systematics.

The species-uniform behavior observed at higher energies motivates a compact
even--odd decomposition,
$\zeta_{\rm bg}=\tfrac12(\zeta_{\rm rf}^{(B)}+\zeta_{\rm rf}^{(\bar B)})$ and
$\Delta\zeta_{\rm rf}=\tfrac12(\zeta_{\rm rf}^{(\bar B)}-\zeta_{\rm rf}^{(B)})$,
which cleanly isolates the charge-odd component of the radial-flow response.
Light nuclei, such as the deuteron ($|n_B|=2$), provide an additional lever arm
through the expected scaling $\zeta_b=(1-\zeta_{\rm rf})^{|n_B|}$.
Empirically, the stability of $\Delta\zeta_{\rm rf}$ under reasonable variations
of the $p_T$ range indicates that the charge-odd signal remains robust over the
beam-energy interval where scaling fidelity is maintained.

Figure~\ref{fig6}(b) summarizes the magnitude of the baryon--antibaryon
separation, $\tfrac12|\Delta\zeta_{\rm rf}|$, plotted versus the baryon chemical
potential $\mu_B$, obtained from standard statistical--thermal parametrizations
of $\mu_B(\sqrt{s_{NN}})$~\cite{Cleymans:2005xv,Andronic:2017pug}.
Over most of the measured range, $\Delta\zeta_{\rm rf}$ evolves smoothly with
$\mu_B$, consistent with the corresponding trends of
$\zeta_{\rm rf}^{(B)}$ and $\zeta_{\rm rf}^{(\bar B)}$ shown in
Fig.~\ref{fig6}(a).
The splitting increases systematically with increasing $\mu_B$ and becomes
clearly nonzero in the high-$\mu_B$ (low-$\sqrt{s_{NN}}$) regime, directly
linking the growth of the charge-odd response to rising midrapidity net-baryon
density.

At the largest $\mu_B$ values, modest deviations from this smooth evolution
appear concurrently with reduced antibaryon scaling fidelity and an increase in
the meson re-scattering parameter $\zeta_{\rm hs}$.
These deviations are most pronounced at low ${\rm KE}_T$ and are consistent
with the growing influence of late-stage hadronic dynamics discussed above.

Separately, the correlated rise of $\zeta_{\rm hs}$ and the non-monotonic
evolution of the attenuation scale $k_\beta$ toward low
$\sqrt{s_{NN}}$ (Fig.~\ref{fig5}) are qualitatively consistent with proximity
to a critical region in a finite, rapidly evolving system.
Notably, finite-size scaling analyses of proxy compressibility and net-baryon
fluctuations identify a critical region in the same $\mu_B$ interval associated
with the minimum of $k_\beta$, providing independent support for this
interpretation~\cite{Lacey:2014wqa,Lacey:2024mnv}.

Taken together, Figs.~5 and~6 show that the non-monotonic evolution of medium
attenuation ($k_\beta$), the concurrent growth of hadronic re-scattering
($\zeta_{\rm hs}$), and the emergence of a species-uniform, charge-odd baryon
response ($\Delta\zeta_{\rm rf}$) are dynamically linked as net-baryon transport
increases toward low beam energy.
In this picture, baryon transport not only modifies the bulk medium response but
can also enhance the experimental visibility of critical dynamics in a finite,
rapidly expanding system through correlated scaling observables.

In summary, species-resolved anisotropy scaling functions are extracted from
$v_2(p_T)$ measurements in Pb+Pb and Au+Au collisions across a broad range of
beam energies to isolate medium response in a fully data-driven manner. The
analysis separates viscous attenuation ($k_{\beta}\!\propto\!\eta/s$),
hadronic re-scattering ($\zeta_{\rm hs}$), and a charge-odd radial-flow
blue-shift component that differentiates baryons from antibaryons. Mesons show
a steadily decreasing $\zeta_{\rm hs}$ with increasing $\sqrt{s_{NN}}$, while
$k_{\beta}$ exhibits a non-monotonic beam-energy dependence that correlates
with the low-energy rise of hadronic re-scattering, consistent with
expectations for $\eta/s(T)$ in the vicinity of the
QCD phase transition. For baryons, the radial-flow response evolves smoothly
with $\sqrt{s_{NN}}$, but a species-uniform, charge-odd separation between
baryons and antibaryons emerges toward lower beam energy. The baryon-number
scaling of this separation, including for light nuclei, supports an
interpretation in terms of junction-mediated net-baryon transport at finite
$\mu_B$. At the lowest beam energies, emerging departures from scaling
fidelity—most apparent for antibaryons—indicate that additional late-stage
hadronic or transport-driven effects may contribute on top of the junction
signal. Taken together, the correlated beam-energy evolution of
$k_{\beta}$, $\zeta_{\rm hs}$, and $\Delta\zeta_{\rm rf}$ demonstrates that
species-resolved scaling functions expose linked modifications of medium
opacity, hadronic dynamics, and baryon transport as the system evolves toward
high $\mu_B$.

%
\bibliography{pid-refs-BES}

\providecommand{\noopsort}[1]{}\providecommand{\singleletter}[1]{#1}%
\begin{thebibliography}{48}%
\makeatletter
\providecommand \@ifxundefined [1]{%
 \@ifx{#1\undefined}
}%
\providecommand \@ifnum [1]{%
 \ifnum #1\expandafter \@firstoftwo
 \else \expandafter \@secondoftwo
 \fi
}%
\providecommand \@ifx [1]{%
 \ifx #1\expandafter \@firstoftwo
 \else \expandafter \@secondoftwo
 \fi
}%
\providecommand \natexlab [1]{#1}%
\providecommand \enquote  [1]{``#1''}%
\providecommand \bibnamefont  [1]{#1}%
\providecommand \bibfnamefont [1]{#1}%
\providecommand \citenamefont [1]{#1}%
\providecommand \href@noop [0]{\@secondoftwo}%
\providecommand \href [0]{\begingroup \@sanitize@url \@href}%
\providecommand \@href[1]{\@@startlink{#1}\@@href}%
\providecommand \@@href[1]{\endgroup#1\@@endlink}%
\providecommand \@sanitize@url [0]{\catcode `\\12\catcode `\$12\catcode
  `\&12\catcode `\#12\catcode `\^12\catcode `\_12\catcode `\%12\relax}%
\providecommand \@@startlink[1]{}%
\providecommand \@@endlink[0]{}%
\providecommand \url  [0]{\begingroup\@sanitize@url \@url }%
\providecommand \@url [1]{\endgroup\@href {#1}{\urlprefix }}%
\providecommand \urlprefix  [0]{URL }%
\providecommand \Eprint [0]{\href }%
\providecommand \doibase [0]{http://dx.doi.org/}%
\providecommand \selectlanguage [0]{\@gobble}%
\providecommand \bibinfo  [0]{\@secondoftwo}%
\providecommand \bibfield  [0]{\@secondoftwo}%
\providecommand \translation [1]{[#1]}%
\providecommand \BibitemOpen [0]{}%
\providecommand \bibitemStop [0]{}%
\providecommand \bibitemNoStop [0]{.\EOS\space}%
\providecommand \EOS [0]{\spacefactor3000\relax}%
\providecommand \BibitemShut  [1]{\csname bibitem#1\endcsname}%
\let\auto@bib@innerbib\@empty
\bibitem [{\citenamefont {Ollitrault}(1992)}]{Ollitrault:1992bk}%
  \BibitemOpen
  \bibfield  {author} {\bibinfo {author} {\bibfnamefont {J.-Y.}\ \bibnamefont
  {Ollitrault}},\ }\href {\doibase 10.1103/PhysRevD.46.229} {\bibfield
  {journal} {\bibinfo  {journal} {Phys. Rev.}\ }\textbf {\bibinfo {volume}
  {D46}},\ \bibinfo {pages} {229} (\bibinfo {year} {1992})}\BibitemShut
  {NoStop}%
\bibitem [{\citenamefont {Voloshin}\ \emph {et~al.}(2010)\citenamefont
  {Voloshin}, \citenamefont {Poskanzer},\ and\ \citenamefont
  {Snellings}}]{Voloshin:2008dg}%
  \BibitemOpen
  \bibfield  {author} {\bibinfo {author} {\bibfnamefont {S.~A.}\ \bibnamefont
  {Voloshin}}, \bibinfo {author} {\bibfnamefont {A.~M.}\ \bibnamefont
  {Poskanzer}}, \ and\ \bibinfo {author} {\bibfnamefont {R.}~\bibnamefont
  {Snellings}},\ }\href {\doibase 10.1007/978-3-642-01539-7_10} {\bibfield
  {journal} {\bibinfo  {journal} {Landolt-Bornstein}\ }\textbf {\bibinfo
  {volume} {23}},\ \bibinfo {pages} {293} (\bibinfo {year} {2010})},\ \Eprint
  {http://arxiv.org/abs/0809.2949} {arXiv:0809.2949 [nucl-ex]} \BibitemShut
  {NoStop}%
\bibitem [{\citenamefont {Heinz}\ and\ \citenamefont
  {Snellings}(2013)}]{Heinz:2013th}%
  \BibitemOpen
  \bibfield  {author} {\bibinfo {author} {\bibfnamefont {U.}~\bibnamefont
  {Heinz}}\ and\ \bibinfo {author} {\bibfnamefont {R.}~\bibnamefont
  {Snellings}},\ }\href {\doibase 10.1146/annurev-nucl-102212-170540}
  {\bibfield  {journal} {\bibinfo  {journal} {Ann. Rev. Nucl. Part. Sci.}\
  }\textbf {\bibinfo {volume} {63}},\ \bibinfo {pages} {123} (\bibinfo {year}
  {2013})}\BibitemShut {NoStop}%
\bibitem [{\citenamefont {Gale}\ \emph {et~al.}(2013)\citenamefont {Gale},
  \citenamefont {Jeon},\ and\ \citenamefont {Schenke}}]{Gale:2013da}%
  \BibitemOpen
  \bibfield  {author} {\bibinfo {author} {\bibfnamefont {C.}~\bibnamefont
  {Gale}}, \bibinfo {author} {\bibfnamefont {S.}~\bibnamefont {Jeon}}, \ and\
  \bibinfo {author} {\bibfnamefont {B.}~\bibnamefont {Schenke}},\ }\href@noop
  {} {\bibfield  {journal} {\bibinfo  {journal} {Int. J. Mod. Phys.}\ }\textbf
  {\bibinfo {volume} {A28}},\ \bibinfo {pages} {1340011} (\bibinfo {year}
  {2013})}\BibitemShut {NoStop}%
\bibitem [{\citenamefont {Bjorken}(1982)}]{Bjorken:1982tu}%
  \BibitemOpen
  \bibfield  {author} {\bibinfo {author} {\bibfnamefont {J.~D.}\ \bibnamefont
  {Bjorken}},\ }\href@noop {} {\  (\bibinfo {year} {1982})}\BibitemShut
  {NoStop}%
\bibitem [{\citenamefont {Baier}\ \emph {et~al.}(1997)\citenamefont {Baier},
  \citenamefont {Dokshitzer}, \citenamefont {Mueller}, \citenamefont {Peigne},\
  and\ \citenamefont {Schiff}}]{Baier:1996kr}%
  \BibitemOpen
  \bibfield  {author} {\bibinfo {author} {\bibfnamefont {R.}~\bibnamefont
  {Baier}}, \bibinfo {author} {\bibfnamefont {Y.~L.}\ \bibnamefont
  {Dokshitzer}}, \bibinfo {author} {\bibfnamefont {A.~H.}\ \bibnamefont
  {Mueller}}, \bibinfo {author} {\bibfnamefont {S.}~\bibnamefont {Peigne}}, \
  and\ \bibinfo {author} {\bibfnamefont {D.}~\bibnamefont {Schiff}},\ }\href
  {\doibase 10.1016/S0550-3213(96)00553-6} {\bibfield  {journal} {\bibinfo
  {journal} {Nucl. Phys. B}\ }\textbf {\bibinfo {volume} {483}},\ \bibinfo
  {pages} {291} (\bibinfo {year} {1997})},\ \Eprint
  {http://arxiv.org/abs/hep-ph/9607355} {arXiv:hep-ph/9607355} \BibitemShut
  {NoStop}%
\bibitem [{\citenamefont {Majumder}\ and\ \citenamefont
  {Shen}(2012)}]{Majumder:2011uk}%
  \BibitemOpen
  \bibfield  {author} {\bibinfo {author} {\bibfnamefont {A.}~\bibnamefont
  {Majumder}}\ and\ \bibinfo {author} {\bibfnamefont {C.}~\bibnamefont
  {Shen}},\ }\href {\doibase 10.1103/PhysRevLett.109.202301} {\bibfield
  {journal} {\bibinfo  {journal} {Phys. Rev. Lett.}\ }\textbf {\bibinfo
  {volume} {109}},\ \bibinfo {pages} {202301} (\bibinfo {year} {2012})},\
  \Eprint {http://arxiv.org/abs/1103.0809} {arXiv:1103.0809 [hep-ph]}
  \BibitemShut {NoStop}%
\bibitem [{\citenamefont {Majumder}\ \emph {et~al.}(2007)\citenamefont
  {Majumder}, \citenamefont {Muller},\ and\ \citenamefont
  {Wang}}]{Majumder:2007zh}%
  \BibitemOpen
  \bibfield  {author} {\bibinfo {author} {\bibfnamefont {A.}~\bibnamefont
  {Majumder}}, \bibinfo {author} {\bibfnamefont {B.}~\bibnamefont {Muller}}, \
  and\ \bibinfo {author} {\bibfnamefont {X.-N.}\ \bibnamefont {Wang}},\ }\href
  {\doibase 10.1103/PhysRevLett.99.192301} {\bibfield  {journal} {\bibinfo
  {journal} {Phys. Rev. Lett.}\ }\textbf {\bibinfo {volume} {99}},\ \bibinfo
  {pages} {192301} (\bibinfo {year} {2007})},\ \Eprint
  {http://arxiv.org/abs/hep-ph/0703082} {arXiv:hep-ph/0703082} \BibitemShut
  {NoStop}%
\bibitem [{\citenamefont {Dusling}\ \emph {et~al.}(2010)\citenamefont
  {Dusling}, \citenamefont {Moore},\ and\ \citenamefont
  {Teaney}}]{Dusling:2009df}%
  \BibitemOpen
  \bibfield  {author} {\bibinfo {author} {\bibfnamefont {K.}~\bibnamefont
  {Dusling}}, \bibinfo {author} {\bibfnamefont {G.~D.}\ \bibnamefont {Moore}},
  \ and\ \bibinfo {author} {\bibfnamefont {D.}~\bibnamefont {Teaney}},\ }\href
  {\doibase 10.1103/PhysRevC.81.034907} {\bibfield  {journal} {\bibinfo
  {journal} {Phys.Rev.}\ }\textbf {\bibinfo {volume} {C81}},\ \bibinfo {pages}
  {034907} (\bibinfo {year} {2010})},\ \Eprint {http://arxiv.org/abs/0909.0754}
  {arXiv:0909.0754 [nucl-th]} \BibitemShut {NoStop}%
\bibitem [{\citenamefont {Lacey}(2024{\natexlab{a}})}]{Lacey:2024fpb}%
  \BibitemOpen
  \bibfield  {author} {\bibinfo {author} {\bibfnamefont {R.~A.}\ \bibnamefont
  {Lacey}},\ }\href {\doibase 10.1103/PhysRevC.110.L031901} {\bibfield
  {journal} {\bibinfo  {journal} {Phys. Rev. C}\ }\textbf {\bibinfo {volume}
  {110}},\ \bibinfo {pages} {L031901} (\bibinfo {year} {2024}{\natexlab{a}})},\
  \Eprint {http://arxiv.org/abs/2402.09389} {arXiv:2402.09389 [nucl-ex]}
  \BibitemShut {NoStop}%
\bibitem [{\citenamefont {Lacey}(2024{\natexlab{b}})}]{Lacey:2024uky}%
  \BibitemOpen
  \bibfield  {author} {\bibinfo {author} {\bibfnamefont {R.~A.}\ \bibnamefont
  {Lacey}},\ }\href@noop {} {\  (\bibinfo {year} {2024}{\natexlab{b}})},\
  \Eprint {http://arxiv.org/abs/2410.04329} {arXiv:2410.04329 [nucl-ex]}
  \BibitemShut {NoStop}%
\bibitem [{\citenamefont {Kharzeev}(1996)}]{Kharzeev:1996sq}%
  \BibitemOpen
  \bibfield  {author} {\bibinfo {author} {\bibfnamefont {D.}~\bibnamefont
  {Kharzeev}},\ }\href {\doibase 10.1016/0370-2693(96)00435-2} {\bibfield
  {journal} {\bibinfo  {journal} {Phys. Lett. B}\ }\textbf {\bibinfo {volume}
  {378}},\ \bibinfo {pages} {238} (\bibinfo {year} {1996})},\ \Eprint
  {http://arxiv.org/abs/nucl-th/9602027} {arXiv:nucl-th/9602027} \BibitemShut
  {NoStop}%
\bibitem [{\citenamefont {Rossi}\ and\ \citenamefont
  {Veneziano}(1977)}]{Rossi:1977cy}%
  \BibitemOpen
  \bibfield  {author} {\bibinfo {author} {\bibfnamefont {G.~C.}\ \bibnamefont
  {Rossi}}\ and\ \bibinfo {author} {\bibfnamefont {G.}~\bibnamefont
  {Veneziano}},\ }\href {\doibase 10.1016/0550-3213(77)90178-X} {\bibfield
  {journal} {\bibinfo  {journal} {Nucl. Phys. B}\ }\textbf {\bibinfo {volume}
  {123}},\ \bibinfo {pages} {507} (\bibinfo {year} {1977})}\BibitemShut
  {NoStop}%
\bibitem [{\citenamefont {Vance}\ \emph {et~al.}(1998)\citenamefont {Vance},
  \citenamefont {Gyulassy},\ and\ \citenamefont {Wang}}]{Vance:1998vh}%
  \BibitemOpen
  \bibfield  {author} {\bibinfo {author} {\bibfnamefont {S.~E.}\ \bibnamefont
  {Vance}}, \bibinfo {author} {\bibfnamefont {M.}~\bibnamefont {Gyulassy}}, \
  and\ \bibinfo {author} {\bibfnamefont {X.~N.}\ \bibnamefont {Wang}},\ }\href
  {\doibase 10.1016/S0370-2693(98)01338-0} {\bibfield  {journal} {\bibinfo
  {journal} {Phys. Lett. B}\ }\textbf {\bibinfo {volume} {443}},\ \bibinfo
  {pages} {45} (\bibinfo {year} {1998})},\ \Eprint
  {http://arxiv.org/abs/nucl-th/9806008} {arXiv:nucl-th/9806008} \BibitemShut
  {NoStop}%
\bibitem [{\citenamefont {Lewis}\ \emph {et~al.}(2024)\citenamefont {Lewis},
  \citenamefont {Lv}, \citenamefont {Ross}, \citenamefont {Tsang},
  \citenamefont {Brandenburg}, \citenamefont {Lin}, \citenamefont {Ma},
  \citenamefont {Tang}, \citenamefont {Tribedy},\ and\ \citenamefont
  {Xu}}]{Lewis:2022arg}%
  \BibitemOpen
  \bibfield  {author} {\bibinfo {author} {\bibfnamefont {N.}~\bibnamefont
  {Lewis}}, \bibinfo {author} {\bibfnamefont {W.}~\bibnamefont {Lv}}, \bibinfo
  {author} {\bibfnamefont {M.~A.}\ \bibnamefont {Ross}}, \bibinfo {author}
  {\bibfnamefont {C.~Y.}\ \bibnamefont {Tsang}}, \bibinfo {author}
  {\bibfnamefont {J.~D.}\ \bibnamefont {Brandenburg}}, \bibinfo {author}
  {\bibfnamefont {Z.-W.}\ \bibnamefont {Lin}}, \bibinfo {author} {\bibfnamefont
  {R.}~\bibnamefont {Ma}}, \bibinfo {author} {\bibfnamefont {Z.}~\bibnamefont
  {Tang}}, \bibinfo {author} {\bibfnamefont {P.}~\bibnamefont {Tribedy}}, \
  and\ \bibinfo {author} {\bibfnamefont {Z.}~\bibnamefont {Xu}},\ }\href
  {\doibase 10.1140/epjc/s10052-024-12834-2} {\bibfield  {journal} {\bibinfo
  {journal} {Eur. Phys. J. C}\ }\textbf {\bibinfo {volume} {84}},\ \bibinfo
  {pages} {590} (\bibinfo {year} {2024})},\ \Eprint
  {http://arxiv.org/abs/2205.05685} {arXiv:2205.05685 [hep-ph]} \BibitemShut
  {NoStop}%
\bibitem [{\citenamefont {Magdy}\ \emph {et~al.}(2024)\citenamefont {Magdy},
  \citenamefont {Deshpande}, \citenamefont {Lacey}, \citenamefont {Li},
  \citenamefont {Tribedy},\ and\ \citenamefont {Xu}}]{Magdy:2024dpm}%
  \BibitemOpen
  \bibfield  {author} {\bibinfo {author} {\bibfnamefont {N.}~\bibnamefont
  {Magdy}}, \bibinfo {author} {\bibfnamefont {A.}~\bibnamefont {Deshpande}},
  \bibinfo {author} {\bibfnamefont {R.}~\bibnamefont {Lacey}}, \bibinfo
  {author} {\bibfnamefont {W.}~\bibnamefont {Li}}, \bibinfo {author}
  {\bibfnamefont {P.}~\bibnamefont {Tribedy}}, \ and\ \bibinfo {author}
  {\bibfnamefont {Z.}~\bibnamefont {Xu}},\ }\href@noop {} {\  (\bibinfo {year}
  {2024})},\ \Eprint {http://arxiv.org/abs/2408.07131} {arXiv:2408.07131
  [hep-ph]} \BibitemShut {NoStop}%
\bibitem [{\citenamefont {Pratt}(2024)}]{Pratt:2023pee}%
  \BibitemOpen
  \bibfield  {author} {\bibinfo {author} {\bibfnamefont {S.}~\bibnamefont
  {Pratt}},\ }\href {\doibase 10.1103/PhysRevC.109.044910} {\bibfield
  {journal} {\bibinfo  {journal} {Phys. Rev. C}\ }\textbf {\bibinfo {volume}
  {109}},\ \bibinfo {pages} {044910} (\bibinfo {year} {2024})},\ \Eprint
  {http://arxiv.org/abs/2311.17906} {arXiv:2311.17906 [hep-ph]} \BibitemShut
  {NoStop}%
\bibitem [{\citenamefont {Steinheimer}\ \emph {et~al.}(2012)\citenamefont
  {Steinheimer}, \citenamefont {Koch},\ and\ \citenamefont
  {Bleicher}}]{Steinheimer:2012bn}%
  \BibitemOpen
  \bibfield  {author} {\bibinfo {author} {\bibfnamefont {J.}~\bibnamefont
  {Steinheimer}}, \bibinfo {author} {\bibfnamefont {V.}~\bibnamefont {Koch}}, \
  and\ \bibinfo {author} {\bibfnamefont {M.}~\bibnamefont {Bleicher}},\ }\href
  {\doibase 10.1103/PhysRevC.86.044903} {\bibfield  {journal} {\bibinfo
  {journal} {Phys. Rev. C}\ }\textbf {\bibinfo {volume} {86}},\ \bibinfo
  {pages} {044903} (\bibinfo {year} {2012})},\ \Eprint
  {http://arxiv.org/abs/1207.2791} {arXiv:1207.2791 [nucl-th]} \BibitemShut
  {NoStop}%
\bibitem [{\citenamefont {Sun}\ \emph {et~al.}(2017)\citenamefont {Sun},
  \citenamefont {Chen}, \citenamefont {Ko},\ and\ \citenamefont
  {Xu}}]{Sun:2017xrx}%
  \BibitemOpen
  \bibfield  {author} {\bibinfo {author} {\bibfnamefont {K.-J.}\ \bibnamefont
  {Sun}}, \bibinfo {author} {\bibfnamefont {L.-W.}\ \bibnamefont {Chen}},
  \bibinfo {author} {\bibfnamefont {C.~M.}\ \bibnamefont {Ko}}, \ and\ \bibinfo
  {author} {\bibfnamefont {Z.}~\bibnamefont {Xu}},\ }\href {\doibase
  10.1016/j.physletb.2017.09.056} {\bibfield  {journal} {\bibinfo  {journal}
  {Phys. Lett. B}\ }\textbf {\bibinfo {volume} {774}},\ \bibinfo {pages} {103}
  (\bibinfo {year} {2017})},\ \Eprint {http://arxiv.org/abs/1702.07620}
  {arXiv:1702.07620 [nucl-th]} \BibitemShut {NoStop}%
\bibitem [{\citenamefont {Xu}\ \emph {et~al.}(2012)\citenamefont {Xu},
  \citenamefont {Chen}, \citenamefont {Ko},\ and\ \citenamefont
  {Lin}}]{Xu:2012gf}%
  \BibitemOpen
  \bibfield  {author} {\bibinfo {author} {\bibfnamefont {J.}~\bibnamefont
  {Xu}}, \bibinfo {author} {\bibfnamefont {L.-W.}\ \bibnamefont {Chen}},
  \bibinfo {author} {\bibfnamefont {C.~M.}\ \bibnamefont {Ko}}, \ and\ \bibinfo
  {author} {\bibfnamefont {Z.-W.}\ \bibnamefont {Lin}},\ }\href {\doibase
  10.1103/PhysRevC.85.041901} {\bibfield  {journal} {\bibinfo  {journal} {Phys.
  Rev. C}\ }\textbf {\bibinfo {volume} {85}},\ \bibinfo {pages} {041901}
  (\bibinfo {year} {2012})},\ \Eprint {http://arxiv.org/abs/1201.3391}
  {arXiv:1201.3391 [nucl-th]} \BibitemShut {NoStop}%
\bibitem [{\citenamefont {Zhou}\ and\ \citenamefont
  {Shi}(2025)}]{Zhou:2024cte}%
  \BibitemOpen
  \bibfield  {author} {\bibinfo {author} {\bibfnamefont {S.}~\bibnamefont
  {Zhou}}\ and\ \bibinfo {author} {\bibfnamefont {S.}~\bibnamefont {Shi}},\
  }\href {\doibase 10.1088/0256-307X/42/2/021201} {\bibfield  {journal}
  {\bibinfo  {journal} {Chin. Phys. Lett.}\ }\textbf {\bibinfo {volume} {42}},\
  \bibinfo {pages} {021201} (\bibinfo {year} {2025})},\ \Eprint
  {http://arxiv.org/abs/2410.20765} {arXiv:2410.20765 [nucl-th]} \BibitemShut
  {NoStop}%
\bibitem [{\citenamefont {Zhu}(2019)}]{Zhu:2019twz}%
  \BibitemOpen
  \bibfield  {author} {\bibinfo {author} {\bibfnamefont {Y.}~\bibnamefont
  {Zhu}} (\bibinfo {collaboration} {ALICE}),\ }\href {\doibase
  10.22323/1.340.0441} {\bibfield  {journal} {\bibinfo  {journal} {PoS}\
  }\textbf {\bibinfo {volume} {ICHEP2018}},\ \bibinfo {pages} {441} (\bibinfo
  {year} {2019})}\BibitemShut {NoStop}%
\bibitem [{\citenamefont {Acharya}\ \emph {et~al.}(2020)\citenamefont {Acharya}
  \emph {et~al.}}]{ALICE:2020chv}%
  \BibitemOpen
  \bibfield  {author} {\bibinfo {author} {\bibfnamefont {S.}~\bibnamefont
  {Acharya}} \emph {et~al.} (\bibinfo {collaboration} {ALICE}),\ }\href
  {\doibase 10.1103/PhysRevC.102.055203} {\bibfield  {journal} {\bibinfo
  {journal} {Phys. Rev. C}\ }\textbf {\bibinfo {volume} {102}},\ \bibinfo
  {pages} {055203} (\bibinfo {year} {2020})},\ \Eprint
  {http://arxiv.org/abs/2005.14639} {arXiv:2005.14639 [nucl-ex]} \BibitemShut
  {NoStop}%
\bibitem [{\citenamefont {Acharya}\ \emph {et~al.}(2023)\citenamefont {Acharya}
  \emph {et~al.}}]{ALICE:2022zks}%
  \BibitemOpen
  \bibfield  {author} {\bibinfo {author} {\bibfnamefont {S.}~\bibnamefont
  {Acharya}} \emph {et~al.} (\bibinfo {collaboration} {ALICE}),\ }\href
  {\doibase 10.1007/JHEP05(2023)243} {\bibfield  {journal} {\bibinfo  {journal}
  {JHEP}\ }\textbf {\bibinfo {volume} {05}},\ \bibinfo {pages} {243} (\bibinfo
  {year} {2023})},\ \Eprint {http://arxiv.org/abs/2206.04587} {arXiv:2206.04587
  [nucl-ex]} \BibitemShut {NoStop}%
\bibitem [{\citenamefont {Liu}\ and\ \citenamefont
  {Lacey}(2018)}]{Liu:2018hjh}%
  \BibitemOpen
  \bibfield  {author} {\bibinfo {author} {\bibfnamefont {P.}~\bibnamefont
  {Liu}}\ and\ \bibinfo {author} {\bibfnamefont {R.~A.}\ \bibnamefont
  {Lacey}},\ }\href {\doibase 10.1103/PhysRevC.98.021902} {\bibfield  {journal}
  {\bibinfo  {journal} {Phys. Rev. C}\ }\textbf {\bibinfo {volume} {98}},\
  \bibinfo {pages} {021902} (\bibinfo {year} {2018})},\ \Eprint
  {http://arxiv.org/abs/1802.06595} {arXiv:1802.06595 [nucl-ex]} \BibitemShut
  {NoStop}%
\bibitem [{\citenamefont {Aamodt}\ \emph {et~al.}(2011)\citenamefont {Aamodt}
  \emph {et~al.}}]{ALICE:2010mlf}%
  \BibitemOpen
  \bibfield  {author} {\bibinfo {author} {\bibfnamefont {K.}~\bibnamefont
  {Aamodt}} \emph {et~al.} (\bibinfo {collaboration} {ALICE}),\ }\href
  {\doibase 10.1103/PhysRevLett.106.032301} {\bibfield  {journal} {\bibinfo
  {journal} {Phys. Rev. Lett.}\ }\textbf {\bibinfo {volume} {106}},\ \bibinfo
  {pages} {032301} (\bibinfo {year} {2011})},\ \Eprint
  {http://arxiv.org/abs/1012.1657} {arXiv:1012.1657 [nucl-ex]} \BibitemShut
  {NoStop}%
\bibitem [{\citenamefont {Adam}\ \emph
  {et~al.}(2016{\natexlab{a}})\citenamefont {Adam} \emph
  {et~al.}}]{ALICE:2015juo}%
  \BibitemOpen
  \bibfield  {author} {\bibinfo {author} {\bibfnamefont {J.}~\bibnamefont
  {Adam}} \emph {et~al.} (\bibinfo {collaboration} {ALICE}),\ }\href {\doibase
  10.1103/PhysRevLett.116.222302} {\bibfield  {journal} {\bibinfo  {journal}
  {Phys. Rev. Lett.}\ }\textbf {\bibinfo {volume} {116}},\ \bibinfo {pages}
  {222302} (\bibinfo {year} {2016}{\natexlab{a}})},\ \Eprint
  {http://arxiv.org/abs/1512.06104} {arXiv:1512.06104 [nucl-ex]} \BibitemShut
  {NoStop}%
\bibitem [{\citenamefont {Acharya}\ \emph {et~al.}(2019)\citenamefont {Acharya}
  \emph {et~al.}}]{ALICE:2018cpu}%
  \BibitemOpen
  \bibfield  {author} {\bibinfo {author} {\bibfnamefont {S.}~\bibnamefont
  {Acharya}} \emph {et~al.} (\bibinfo {collaboration} {ALICE}),\ }\href
  {\doibase 10.1016/j.physletb.2018.12.048} {\bibfield  {journal} {\bibinfo
  {journal} {Phys. Lett. B}\ }\textbf {\bibinfo {volume} {790}},\ \bibinfo
  {pages} {35} (\bibinfo {year} {2019})},\ \Eprint
  {http://arxiv.org/abs/1805.04432} {arXiv:1805.04432 [nucl-ex]} \BibitemShut
  {NoStop}%
\bibitem [{\citenamefont {Sirunyan}\ \emph {et~al.}(2019)\citenamefont
  {Sirunyan} \emph {et~al.}}]{CMS:2019gzk}%
  \BibitemOpen
  \bibfield  {author} {\bibinfo {author} {\bibfnamefont {A.~M.}\ \bibnamefont
  {Sirunyan}} \emph {et~al.} (\bibinfo {collaboration} {CMS}),\ }\href
  {\doibase 10.1016/j.physletb.2019.135049} {\bibfield  {journal} {\bibinfo
  {journal} {Phys. Lett. B}\ }\textbf {\bibinfo {volume} {799}},\ \bibinfo
  {pages} {135049} (\bibinfo {year} {2019})},\ \Eprint
  {http://arxiv.org/abs/1902.03603} {arXiv:1902.03603 [hep-ex]} \BibitemShut
  {NoStop}%
\bibitem [{\citenamefont {Lacey}\ \emph {et~al.}(2016)\citenamefont {Lacey},
  \citenamefont {Liu}, \citenamefont {Magdy}, \citenamefont {Csanád},
  \citenamefont {Schweid}, \citenamefont {Ajitanand}, \citenamefont
  {Alexander},\ and\ \citenamefont {Pak}}]{Lacey:2016hqy}%
  \BibitemOpen
  \bibfield  {author} {\bibinfo {author} {\bibfnamefont {R.~A.}\ \bibnamefont
  {Lacey}}, \bibinfo {author} {\bibfnamefont {P.}~\bibnamefont {Liu}}, \bibinfo
  {author} {\bibfnamefont {N.}~\bibnamefont {Magdy}}, \bibinfo {author}
  {\bibfnamefont {M.}~\bibnamefont {Csanád}}, \bibinfo {author} {\bibfnamefont
  {B.}~\bibnamefont {Schweid}}, \bibinfo {author} {\bibfnamefont {N.~N.}\
  \bibnamefont {Ajitanand}}, \bibinfo {author} {\bibfnamefont {J.}~\bibnamefont
  {Alexander}}, \ and\ \bibinfo {author} {\bibfnamefont {R.}~\bibnamefont
  {Pak}},\ }\href@noop {} {\  (\bibinfo {year} {2016})},\ \Eprint
  {http://arxiv.org/abs/1601.06001} {arXiv:1601.06001 [nucl-ex]} \BibitemShut
  {NoStop}%
\bibitem [{\citenamefont {Miller}\ \emph {et~al.}(2007)\citenamefont {Miller},
  \citenamefont {Reygers}, \citenamefont {Sanders},\ and\ \citenamefont
  {Steinberg}}]{Miller:2007ri}%
  \BibitemOpen
  \bibfield  {author} {\bibinfo {author} {\bibfnamefont {M.~L.}\ \bibnamefont
  {Miller}}, \bibinfo {author} {\bibfnamefont {K.}~\bibnamefont {Reygers}},
  \bibinfo {author} {\bibfnamefont {S.~J.}\ \bibnamefont {Sanders}}, \ and\
  \bibinfo {author} {\bibfnamefont {P.}~\bibnamefont {Steinberg}},\ }\href
  {\doibase 10.1146/annurev.nucl.57.090506.123020} {\bibfield  {journal}
  {\bibinfo  {journal} {Ann. Rev. Nucl. Part. Sci.}\ }\textbf {\bibinfo
  {volume} {57}},\ \bibinfo {pages} {205} (\bibinfo {year} {2007})},\ \Eprint
  {http://arxiv.org/abs/nucl-ex/0701025} {arXiv:nucl-ex/0701025} \BibitemShut
  {NoStop}%
\bibitem [{\citenamefont {Alver}\ \emph {et~al.}(2007)\citenamefont {Alver}
  \emph {et~al.}}]{PHOBOS:2006dbo}%
  \BibitemOpen
  \bibfield  {author} {\bibinfo {author} {\bibfnamefont {B.}~\bibnamefont
  {Alver}} \emph {et~al.} (\bibinfo {collaboration} {PHOBOS}),\ }\href
  {\doibase 10.1103/PhysRevLett.98.242302} {\bibfield  {journal} {\bibinfo
  {journal} {Phys. Rev. Lett.}\ }\textbf {\bibinfo {volume} {98}},\ \bibinfo
  {pages} {242302} (\bibinfo {year} {2007})},\ \Eprint
  {http://arxiv.org/abs/nucl-ex/0610037} {arXiv:nucl-ex/0610037} \BibitemShut
  {NoStop}%
\bibitem [{\citenamefont {Abelev}\ \emph {et~al.}(2015)\citenamefont {Abelev}
  \emph {et~al.}}]{ALICE:2014wao}%
  \BibitemOpen
  \bibfield  {author} {\bibinfo {author} {\bibfnamefont {B.~B.}\ \bibnamefont
  {Abelev}} \emph {et~al.} (\bibinfo {collaboration} {ALICE}),\ }\href
  {\doibase 10.1007/JHEP06(2015)190} {\bibfield  {journal} {\bibinfo  {journal}
  {JHEP}\ }\textbf {\bibinfo {volume} {06}},\ \bibinfo {pages} {190} (\bibinfo
  {year} {2015})},\ \Eprint {http://arxiv.org/abs/1405.4632} {arXiv:1405.4632
  [nucl-ex]} \BibitemShut {NoStop}%
\bibitem [{\citenamefont {Adam}\ \emph
  {et~al.}(2016{\natexlab{b}})\citenamefont {Adam} \emph
  {et~al.}}]{ALICE:2016cti}%
  \BibitemOpen
  \bibfield  {author} {\bibinfo {author} {\bibfnamefont {J.}~\bibnamefont
  {Adam}} \emph {et~al.} (\bibinfo {collaboration} {ALICE}),\ }\href {\doibase
  10.1007/JHEP09(2016)164} {\bibfield  {journal} {\bibinfo  {journal} {JHEP}\
  }\textbf {\bibinfo {volume} {09}},\ \bibinfo {pages} {164} (\bibinfo {year}
  {2016}{\natexlab{b}})},\ \Eprint {http://arxiv.org/abs/1606.06057}
  {arXiv:1606.06057 [nucl-ex]} \BibitemShut {NoStop}%
\bibitem [{\citenamefont {Acharya}\ \emph {et~al.}(2017)\citenamefont {Acharya}
  \emph {et~al.}}]{ALICE:2017nuf}%
  \BibitemOpen
  \bibfield  {author} {\bibinfo {author} {\bibfnamefont {S.}~\bibnamefont
  {Acharya}} \emph {et~al.} (\bibinfo {collaboration} {ALICE}),\ }\href
  {\doibase 10.1140/epjc/s10052-017-5222-x} {\bibfield  {journal} {\bibinfo
  {journal} {Eur. Phys. J. C}\ }\textbf {\bibinfo {volume} {77}},\ \bibinfo
  {pages} {658} (\bibinfo {year} {2017})},\ \Eprint
  {http://arxiv.org/abs/1707.07304} {arXiv:1707.07304 [nucl-ex]} \BibitemShut
  {NoStop}%
\bibitem [{\citenamefont {Acharya}\ \emph {et~al.}(2018)\citenamefont {Acharya}
  \emph {et~al.}}]{ALICE:2018lao}%
  \BibitemOpen
  \bibfield  {author} {\bibinfo {author} {\bibfnamefont {S.}~\bibnamefont
  {Acharya}} \emph {et~al.} (\bibinfo {collaboration} {ALICE}),\ }\href
  {\doibase 10.1016/j.physletb.2018.06.059} {\bibfield  {journal} {\bibinfo
  {journal} {Phys. Lett. B}\ }\textbf {\bibinfo {volume} {784}},\ \bibinfo
  {pages} {82} (\bibinfo {year} {2018})},\ \Eprint
  {http://arxiv.org/abs/1805.01832} {arXiv:1805.01832 [nucl-ex]} \BibitemShut
  {NoStop}%
\bibitem [{\citenamefont {Adamczyk}\ \emph {et~al.}(2013)\citenamefont
  {Adamczyk} \emph {et~al.}}]{STAR:2013ayu}%
  \BibitemOpen
  \bibfield  {author} {\bibinfo {author} {\bibfnamefont {L.}~\bibnamefont
  {Adamczyk}} \emph {et~al.} (\bibinfo {collaboration} {STAR}),\ }\href
  {\doibase 10.1103/PhysRevC.88.014902} {\bibfield  {journal} {\bibinfo
  {journal} {Phys. Rev. C}\ }\textbf {\bibinfo {volume} {88}},\ \bibinfo
  {pages} {014902} (\bibinfo {year} {2013})},\ \Eprint
  {http://arxiv.org/abs/1301.2348} {arXiv:1301.2348 [nucl-ex]} \BibitemShut
  {NoStop}%
\bibitem [{\citenamefont {Adare}\ \emph {et~al.}(2016)\citenamefont {Adare}
  \emph {et~al.}}]{PHENIX:2014uik}%
  \BibitemOpen
  \bibfield  {author} {\bibinfo {author} {\bibfnamefont {A.}~\bibnamefont
  {Adare}} \emph {et~al.} (\bibinfo {collaboration} {PHENIX}),\ }\href
  {\doibase 10.1103/PhysRevC.93.051902} {\bibfield  {journal} {\bibinfo
  {journal} {Phys. Rev. C}\ }\textbf {\bibinfo {volume} {93}},\ \bibinfo
  {pages} {051902} (\bibinfo {year} {2016})},\ \Eprint
  {http://arxiv.org/abs/1412.1038} {arXiv:1412.1038 [nucl-ex]} \BibitemShut
  {NoStop}%
\bibitem [{\citenamefont {Adamczyk}\ \emph {et~al.}(2016)\citenamefont
  {Adamczyk} \emph {et~al.}}]{STAR:2015gge}%
  \BibitemOpen
  \bibfield  {author} {\bibinfo {author} {\bibfnamefont {L.}~\bibnamefont
  {Adamczyk}} \emph {et~al.} (\bibinfo {collaboration} {STAR}),\ }\href
  {\doibase 10.1103/PhysRevLett.116.062301} {\bibfield  {journal} {\bibinfo
  {journal} {Phys. Rev. Lett.}\ }\textbf {\bibinfo {volume} {116}},\ \bibinfo
  {pages} {062301} (\bibinfo {year} {2016})},\ \Eprint
  {http://arxiv.org/abs/1507.05247} {arXiv:1507.05247 [nucl-ex]} \BibitemShut
  {NoStop}%
\bibitem [{\citenamefont {Abdallah}\ \emph {et~al.}(2022)\citenamefont
  {Abdallah} \emph {et~al.}}]{STAR:2022ncy}%
  \BibitemOpen
  \bibfield  {author} {\bibinfo {author} {\bibfnamefont {M.}~\bibnamefont
  {Abdallah}} \emph {et~al.} (\bibinfo {collaboration} {STAR}),\ }\href
  {\doibase 10.1103/PhysRevC.105.064911} {\bibfield  {journal} {\bibinfo
  {journal} {Phys. Rev. C}\ }\textbf {\bibinfo {volume} {105}},\ \bibinfo
  {pages} {064911} (\bibinfo {year} {2022})},\ \Eprint
  {http://arxiv.org/abs/2203.07204} {arXiv:2203.07204 [nucl-ex]} \BibitemShut
  {NoStop}%
\bibitem [{\citenamefont {Dokshitzer}\ and\ \citenamefont
  {Kharzeev}(2001)}]{Dokshitzer:2001zm}%
  \BibitemOpen
  \bibfield  {author} {\bibinfo {author} {\bibfnamefont {Y.~L.}\ \bibnamefont
  {Dokshitzer}}\ and\ \bibinfo {author} {\bibfnamefont {D.~E.}\ \bibnamefont
  {Kharzeev}},\ }\href {\doibase 10.1016/S0370-2693(01)01130-3} {\bibfield
  {journal} {\bibinfo  {journal} {Phys. Lett. B}\ }\textbf {\bibinfo {volume}
  {519}},\ \bibinfo {pages} {199} (\bibinfo {year} {2001})},\ \Eprint
  {http://arxiv.org/abs/hep-ph/0106202} {arXiv:hep-ph/0106202} \BibitemShut
  {NoStop}%
\bibitem [{\citenamefont {Lacey}\ \emph {et~al.}(2010)\citenamefont {Lacey},
  \citenamefont {Taranenko}, \citenamefont {Wei}, \citenamefont {Ajitanand},
  \citenamefont {Alexander} \emph {et~al.}}]{Lacey:2010fe}%
  \BibitemOpen
  \bibfield  {author} {\bibinfo {author} {\bibfnamefont {R.~A.}\ \bibnamefont
  {Lacey}}, \bibinfo {author} {\bibfnamefont {A.}~\bibnamefont {Taranenko}},
  \bibinfo {author} {\bibfnamefont {R.}~\bibnamefont {Wei}}, \bibinfo {author}
  {\bibfnamefont {N.}~\bibnamefont {Ajitanand}}, \bibinfo {author}
  {\bibfnamefont {J.}~\bibnamefont {Alexander}},  \emph {et~al.},\ }\href
  {\doibase 10.1103/PhysRevC.82.034910} {\bibfield  {journal} {\bibinfo
  {journal} {Phys.Rev.}\ }\textbf {\bibinfo {volume} {C82}},\ \bibinfo {pages}
  {034910} (\bibinfo {year} {2010})},\ \Eprint {http://arxiv.org/abs/1005.4979}
  {arXiv:1005.4979 [nucl-ex]} \BibitemShut {NoStop}%
\bibitem [{\citenamefont {Csernai}\ \emph {et~al.}(2006)\citenamefont
  {Csernai}, \citenamefont {Kapusta},\ and\ \citenamefont
  {McLerran}}]{Csernai:2006zz}%
  \BibitemOpen
  \bibfield  {author} {\bibinfo {author} {\bibfnamefont {L.~P.}\ \bibnamefont
  {Csernai}}, \bibinfo {author} {\bibfnamefont {J.~I.}\ \bibnamefont
  {Kapusta}}, \ and\ \bibinfo {author} {\bibfnamefont {L.~D.}\ \bibnamefont
  {McLerran}},\ }\href {\doibase 10.1103/PhysRevLett.97.152303} {\bibfield
  {journal} {\bibinfo  {journal} {Phys. Rev. Lett.}\ }\textbf {\bibinfo
  {volume} {97}},\ \bibinfo {pages} {152303} (\bibinfo {year} {2006})},\
  \Eprint {http://arxiv.org/abs/nucl-th/0604032} {arXiv:nucl-th/0604032}
  \BibitemShut {NoStop}%
\bibitem [{\citenamefont {Lacey}\ \emph {et~al.}(2007)\citenamefont {Lacey},
  \citenamefont {Ajitanand}, \citenamefont {Alexander}, \citenamefont {Chung},
  \citenamefont {Holzmann} \emph {et~al.}}]{Lacey:2006bc}%
  \BibitemOpen
  \bibfield  {author} {\bibinfo {author} {\bibfnamefont {R.~A.}\ \bibnamefont
  {Lacey}}, \bibinfo {author} {\bibfnamefont {N.}~\bibnamefont {Ajitanand}},
  \bibinfo {author} {\bibfnamefont {J.}~\bibnamefont {Alexander}}, \bibinfo
  {author} {\bibfnamefont {P.}~\bibnamefont {Chung}}, \bibinfo {author}
  {\bibfnamefont {W.}~\bibnamefont {Holzmann}},  \emph {et~al.},\ }\href
  {\doibase 10.1103/PhysRevLett.98.092301} {\bibfield  {journal} {\bibinfo
  {journal} {Phys.Rev.Lett.}\ }\textbf {\bibinfo {volume} {98}},\ \bibinfo
  {pages} {092301} (\bibinfo {year} {2007})},\ \Eprint
  {http://arxiv.org/abs/nucl-ex/0609025} {arXiv:nucl-ex/0609025 [nucl-ex]}
  \BibitemShut {NoStop}%
\bibitem [{\citenamefont {Cleymans}\ \emph {et~al.}(2006)\citenamefont
  {Cleymans}, \citenamefont {Oeschler}, \citenamefont {Redlich},\ and\
  \citenamefont {Wheaton}}]{Cleymans:2005xv}%
  \BibitemOpen
  \bibfield  {author} {\bibinfo {author} {\bibfnamefont {J.}~\bibnamefont
  {Cleymans}}, \bibinfo {author} {\bibfnamefont {H.}~\bibnamefont {Oeschler}},
  \bibinfo {author} {\bibfnamefont {K.}~\bibnamefont {Redlich}}, \ and\
  \bibinfo {author} {\bibfnamefont {S.}~\bibnamefont {Wheaton}},\ }\href
  {\doibase 10.1103/PhysRevC.73.034905} {\bibfield  {journal} {\bibinfo
  {journal} {Phys. Rev. C}\ }\textbf {\bibinfo {volume} {73}},\ \bibinfo
  {pages} {034905} (\bibinfo {year} {2006})},\ \Eprint
  {http://arxiv.org/abs/hep-ph/0511094} {arXiv:hep-ph/0511094} \BibitemShut
  {NoStop}%
\bibitem [{\citenamefont {Andronic}\ \emph {et~al.}(2018)\citenamefont
  {Andronic}, \citenamefont {Braun-Munzinger}, \citenamefont {Redlich},\ and\
  \citenamefont {Stachel}}]{Andronic:2017pug}%
  \BibitemOpen
  \bibfield  {author} {\bibinfo {author} {\bibfnamefont {A.}~\bibnamefont
  {Andronic}}, \bibinfo {author} {\bibfnamefont {P.}~\bibnamefont
  {Braun-Munzinger}}, \bibinfo {author} {\bibfnamefont {K.}~\bibnamefont
  {Redlich}}, \ and\ \bibinfo {author} {\bibfnamefont {J.}~\bibnamefont
  {Stachel}},\ }\href {\doibase 10.1038/s41586-018-0491-6} {\bibfield
  {journal} {\bibinfo  {journal} {Nature}\ }\textbf {\bibinfo {volume} {561}},\
  \bibinfo {pages} {321} (\bibinfo {year} {2018})},\ \Eprint
  {http://arxiv.org/abs/1710.09425} {arXiv:1710.09425 [nucl-th]} \BibitemShut
  {NoStop}%
\bibitem [{\citenamefont {Lacey}(2015)}]{Lacey:2014wqa}%
  \BibitemOpen
  \bibfield  {author} {\bibinfo {author} {\bibfnamefont {R.~A.}\ \bibnamefont
  {Lacey}},\ }\href {\doibase 10.1103/PhysRevLett.114.142301} {\bibfield
  {journal} {\bibinfo  {journal} {Phys. Rev. Lett.}\ }\textbf {\bibinfo
  {volume} {114}},\ \bibinfo {pages} {142301} (\bibinfo {year} {2015})},\
  \Eprint {http://arxiv.org/abs/1411.7931} {arXiv:1411.7931 [nucl-ex]}
  \BibitemShut {NoStop}%
\bibitem [{\citenamefont {Lacey}(2024{\natexlab{c}})}]{Lacey:2024mnv}%
  \BibitemOpen
  \bibfield  {author} {\bibinfo {author} {\bibfnamefont {R.~A.}\ \bibnamefont
  {Lacey}},\ }\href@noop {} {\  (\bibinfo {year} {2024}{\natexlab{c}})},\
  \Eprint {http://arxiv.org/abs/2411.09139} {arXiv:2411.09139 [nucl-ex]}
  \BibitemShut {NoStop}%
\end{thebibliography}%

\end{document}